\documentclass[pdflatex,sn-mathphys-num]{sn-jnl}

\usepackage{graphicx}%
\usepackage{multirow}%
\usepackage{amsmath,amssymb,amsfonts}%
\usepackage{amsthm}%
\usepackage{mathrsfs}%
\usepackage[title]{appendix}%
\usepackage{xcolor}%
\usepackage{textcomp}%
\usepackage{manyfoot}%
\usepackage{booktabs}%
\usepackage{algorithm}%
\usepackage{algorithmicx}%
\usepackage{algpseudocode}%
\usepackage{listings}%
\usepackage{adjustbox}
\usepackage{framed}
\usepackage{soul}
\usepackage{hyperref}
\usepackage{lscape}
\usepackage{enumitem}
\usepackage{fontawesome}
\usepackage{multirow}
\usepackage[framemethod=tikz]{mdframed}

\usepackage{xspace}

\theoremstyle{thmstyleone}%
\theoremstyle{thmstyletwo}%

\theoremstyle{thmstylethree}%

\newcommand*{\tool}{\texttt{PatchTrack}\xspace}
\newcommand*{\PA}{\texttt{PA}\xspace} 
\newcommand*{\PN}{\texttt{PN}\xspace} 
\newcommand*{\NE}{\texttt{NE}\xspace} 
\newcommand*{\CC}{\texttt{CC}\xspace} 
\newcommand*{\EE}{\texttt{EE}\xspace} 
\newcommand*{\CL}{\texttt{CL}\xspace} 

\newenvironment{custombox}{\smallskip\begin{mdframed}[linewidth=1pt,innerleftmargin=5pt, innerrightmargin=5pt, innertopmargin=5pt, innerbottommargin=5pt, nobreak=true]}{\end{mdframed}\smallskip}

\usepackage{relsize}
\begin{document}

\title[PatchTrack: A Comprehensive Analysis of ChatGPT's Influence on Pull Request Outcomes]{PatchTrack: A Comprehensive Analysis of ChatGPT’s Influence on Pull Request Outcomes}


\author{\fnm{Daniel} \sur{Ogenrwot}}\email{ogenrwot@unlv.nevada.edu}


\author*{\fnm{John} \sur{Businge}}\email{john.businge@unlv.edu}

\affil{\orgdiv{Department of Computer Science}, \orgname{University of Nevada Las Vegas}, \orgaddress{\street{4505 S. Maryland Pkwy.}, \city{Las Vegas}, \postcode{89154}, \state{Nevada}, \country{United States of America}}}




\abstract{
The rapid adoption of large language models (LLMs) like ChatGPT has introduced new dynamics in software development, particularly within pull request workflows. While prior research has examined the quality of AI-generated code, less is known about how developers evaluate, adapt, and integrate these suggestions in real-world collaboration. We analyze 338 pull requests from 255 GitHub repositories containing self-admitted ChatGPT usage, comprising 645 AI-generated snippets and 3,486 developer-authored patches. To support this analysis at scale, we use PatchTrack, an automated classifier that identifies whether AI-generated patches were applied, partially reused, or not integrated. Our findings reveal that full adoption of ChatGPT-generated code is uncommon: the median integration rate is 25\%. Qualitative analysis of 89 pull requests with integrated patches reveals recurring patterns of \textit{structural integration}, \textit{selective extraction}, and \textit{iterative refinement}, indicating that developers typically treat AI output as a starting point rather than a final implementation. Even when code is not directly adopted, ChatGPT influences workflows through \textit{conceptual guidance}, \textit{documentation}, and \textit{debugging strategies}. Integration decisions reflect \textit{contextual fit}, \textit{integration effort}, \textit{maintainer trust}, and established pull request review norms rather than serving as direct indicators of code correctness. Overall, this study provides empirical insight into AI-mediated decision-making in collaborative software development, showing that the influence of generative AI extends beyond patch generation to how developers reason about, adapt, and negotiate code during review within pull request workflows. These findings inform the design of AI-assisted tools and support more transparent and effective use of LLMs in practice.
}




\maketitle

\section{Introduction}\label{sec1}

AI-assisted workflows are increasingly shaping software development, with tools such as ChatGPT and GitHub Copilot that promise improved automation, productivity, and collaboration~\cite{menzies2020software, Russo:generativeAI:2024, huang2024generative, Ebert:GenAI:2023, sauvola2024future}. Prior work has examined their utility in code generation, debugging, and documentation~\cite{ozpolat2023ai_tools, peng2023impact, wermelinger2023using, jin2024can}, but most studies have focused on controlled settings or isolated tasks. Their impact on collaborative workflows, particularly pull request (PR) decision-making, remains poorly understood~\cite{grewal2024analyzing, siddiq2024quality}. Understanding how developers integrate, modify, or reject AI-generated code in real-world PRs is essential to improving AI-assisted collaboration.

Consider a common scenario in open-source development: a contributor submits a pull request and notes that part of the patch was suggested by ChatGPT. During review, maintainers must decide whether the suggested code should be integrated as-is, adapted, or rejected. Objections often concern contextual fit, architectural alignment, or interaction with surrounding code rather than the underlying idea. As a result, contributors may revise the patch, extract only fragments of the suggestion, or abandon it entirely. Such outcomes are typically invisible in analyses that rely solely on merge status or coarse-grained pull request metadata. This study is motivated by the need to empirically characterize these decision-making processes at scale and to examine how AI-generated suggestions influence pull request outcomes even when the code itself is modified or rejected.

To understand why these decision processes matter, it is important to situate them within the broader role of pull requests in modern software engineering. Pull requests enable peer review, discussion, and integration~\cite{Rigby:2013, Bacchelli:2013, storey2016social}. Traditional studies have shown that PR outcomes depend on factors such as patch size, contributor history, and code quality~\cite{Gousios:ICSE:2014, Tsay-2014, Gousios:ICSE:2016, soares2015acceptance, Zhu:PR:2016}. Recently, tools like GitHub Copilot have entered the PR workflow~\cite{xiao2024generative}, offering autogenerated summaries and suggestions to speed up reviews. These advances echo the long-standing vision of the \textit{Programmer’s Apprentice}~\cite{rich1990programmers}, in which AI augments rather than replaces human reasoning. Yet, the effectiveness of such tools in shaping real pull request decisions remains poorly understood.

Although prior research has explored predictive models and recommendation systems to support pull request decision-making~\cite{Zhao:EMSE:2019, Azeem:ICSSP:2020, Dey:ESEM:2020}, these studies typically rely on social and technical metadata such as contributor history or patch size rather than on the content of patches themselves. In parallel, research in AI-assisted software engineering has focused primarily on improving code quality, debugging, documentation, and developer productivity through generative models~\cite{hassan2024rethinking, white2023prompt, luo2024taiyi, howard-ruder-2018-universal, Jiang:2023, Guo:2024, Deng:2024, hou2023large, ju2023llama, siddiq2023exploring}. While these studies demonstrate technical and productivity benefits, often in controlled or task-specific settings, they do not fully examine how AI-generated outputs are judged, adapted, or discarded in collaborative workflows such as pull requests.

Even studies of ChatGPT-generated code that report integration rates and patch retention~\cite{grewal2024analyzing, siddiq2024quality} do not fully capture the nuanced decisions developers make during pull request discussions. To move beyond binary merge outcomes, we therefore examine whether developers apply ChatGPT-suggested code directly, modify it substantially, act only on conceptual guidance, or ultimately reject it, providing a more fine-grained view of AI-mediated decision-making in pull request workflows.

Meanwhile, developers are increasingly incorporating ChatGPT into daily development~\cite{tufano2024chatgpt}, yet recent evaluations of automated code review tools show that such systems often overlook human decision-making context, leading to challenges in accuracy, trust, and integration~\cite{tufano202:IEEEPress}. Hallucinated or incorrect code further undermines adoption, as developers must verify or revise suggestions~\cite{tanzil2024chatgpt}. Multi-hunk patches pose additional challenges for large language models due to limitations in semantic coordination and global reasoning~\cite{nashid2025characterizing}. While these technical and socio-technical barriers are increasingly recognized, few studies have examined them empirically in the context of real-world pull request outcomes.

This study addresses that gap by analyzing 338 pull request conversations from open-source projects involving self-admitted ChatGPT usage (SACU). We introduce and apply \texttt{PatchTrack} as an analytical instrument for classifying patch outcomes and usage patterns under AI assistance, enabling large-scale empirical observation of patch-level integration outcomes rather than providing a developer-facing or prescriptive tool. Originally presented as a poster at ASE 2024, we significantly extend PatchTrack’s capabilities by improving classification precision, validating its outputs, and applying it to a larger, SACU-enhanced version of the DevGPT dataset~\cite{xiao_2023_8304091}. This enables a scalable and systematic analysis of how ChatGPT-generated suggestions are integrated, adapted, or rejected in collaborative workflows.

Unlike prior work that focuses primarily on integration rates or isolated productivity metrics~\cite{Jiang:2023, Guo:2024, Deng:2024, hou2023large, ju2023llama, siddiq2023exploring, li2024copilot}, we adopt a behavioral lens that examines how developers interpret, adapt, or disregard ChatGPT’s outputs within real-world team workflows. Our findings reveal that developers frequently use ChatGPT for purposes beyond code generation, such as conceptual guidance, naming, documentation refinement, and debugging support. We observe a spectrum of use, from direct integration to complete rejection, shaped by contextual judgment, trust, and project norms. This perspective contributes a more grounded understanding of AI-assisted collaboration, informing both tool design and responsible adoption.

\vspace{5pt}
\noindent The paper makes the following contributions:
\begin{itemize}[leftmargin=*] \setlength\itemsep{0em}
    \item We introduce and apply \texttt{PatchTrack} as an analytical instrument to empirically study how developers integrate, adapt, or reject ChatGPT-generated code in real pull requests.
    
    \item We provide a mixed-methods analysis of applied patches, identifying recurring patterns of \textit{structural integration}, \textit{selective extraction}, and \textit{iterative refinement}, showing that developers treat AI-generated code as a starting point rather than a final solution.

    \item We show that ChatGPT influences pull request outcomes even when no code is adopted, through conceptual guidance, documentation refinement, and debugging support.
    
    \item We identify cases where valid AI-generated patches are discarded due to architectural misfit, contributor role, or workflow norms, highlighting socio-technical barriers to adoption.

    \item We release a public replication package~\cite{patchtrack:2025} to support reproducibility and future research on AI-assisted collaboration.
\end{itemize}

\section{Terminology, Problem, and Concrete Examples}
In this section, we define key terminology used in this study, outline the classification of ChatGPT-assisted developer interactions, and present a concrete example to illustrate these concepts.
\subsection{Terminology}
\begin{itemize}[leftmargin=*]
\setlength{\itemsep}{0pt} 
     \item \textbf{Code snippet:} A small, standalone block of code intended for reuse, demonstrating a specific task or programming concept. In this study, ``code snippet'' refers exclusively to discrete pieces of code that ChatGPT suggests, not including code examples embedded within descriptive text or theoretical guidance.

\item \textbf{Diff:} The difference between two versions of files, highlighting the changes between an original and a modified file.

\item \textbf{Patch:} A file containing differences between two versions of one or more files. A patch is a specific type of diff that outlines the changes from one state to another.  

\item \textbf{Hunk:} A contiguous block of changes within a diff or patch, indicating differing lines between two versions of a file~\cite{hunk}. Hunks are identified by headers in the format \texttt{@@ -s,n +s,n @@}, which mark the start and the number of lines affected in the patch.

\item \textbf{Self-admitted ChatGPT usage (SACU):} Instances where developers explicitly acknowledge, within a pull request discussion or issue thread, that they used ChatGPT to generate, modify, or assist in writing code. This admission is typically evident in pull request comments, commit messages, discussion threads, or issue reports where developers refer to AI-generated suggestions.

\item \textbf{PatchTrack: } A classification tool that determines whether ChatGPT-generated patches were applied, not applied, or not suggested during developers' integration with ChatGPT.

\end{itemize}

\subsection{Classifying Developer Interaction with ChatGPT}

A typical GitHub pull request workflow augmented with ChatGPT interactions can originate from the pull request author or a reviewer. Consider a scenario in which a developer, intending to fix a bug, add a feature, or improve an existing one, initiates a change to a file named \texttt{Foo}. Following a pull-based development workflow \cite{Gousios:ICSE:2014}, the developer creates a new branch, commits changes, and eventually merges the pull request into the main repository of the project.

During development, the developer might encounter challenges or seek optimal solutions, prompting them to consult ChatGPT. The responses provided by ChatGPT may be textual explanations or code snippets. In this study, we focus on self-admitted ChatGPT usage (SACU), where developers explicitly acknowledge using ChatGPT within pull request discussions. Based on the self-admitted ChatGPT usage interactions, we classify ChatGPT-assisted pull requests into three possible scenarios:

\begin{enumerate}[leftmargin=*]
\setlength{\itemsep}{0pt} 
    \item When the response includes a code snippet, the developer may apply this directly to \texttt{Foo} as a patch, either as provided or after modifications. This scenario is labeled as ``Patch Applied (\PA)''.
    \item If the code snippets are not applied, or modified beyond recognition (resulting in a semantic clone), then \texttt{Foo} may remain unchanged or include changes influenced by the discussed concepts. This scenario is defined as ``Patch Not Applied (\PN)''.
    \item If the interaction with ChatGPT involves only textual advice without code snippets, similar to the previous scenario, \texttt{Foo} may remain unchanged or may be modified based on the guidance received. This scenario is referred to as ``No Existing Patch (\NE)''.
\end{enumerate}

This categorization underpins our analysis by structuring the diverse ways developers interact with ChatGPT in pull requests. By distinguishing between directly applied patches (\PA), substantially modified or conceptually reused suggestions (\PN), conceptual-only interactions without code (\NE), and rejected contributions in closed pull requests (\CL), we can (i) trace ChatGPT’s concrete influence on code changes, (ii) identify barriers to adoption, trust, and contextual fit, (iii) examine decision-making dynamics in AI-assisted workflows, and (iv) analyze failure cases where AI involvement did not lead to integration. The inclusion of \CL cases is particularly valuable for understanding how project constraints, scope misalignment, or collaboration breakdowns affect AI-supported contributions. This classification also informs targeted recommendations: improving the usability of the patch (\PA), aligning suggestions with developer expectations (\PN), improving conceptual guidance (\NE) and mitigating risks in unmerged PRs (\CL). Together, these categories link observed outcomes to distinct usage patterns, supporting evidence-based insights into how generative AI can enhance the efficiency and reliability of collaborative code review.

Our classification relies on the self-admitted ChatGPT usage, meaning that only cases where developers explicitly state that ChatGPT was used are included in the dataset. This ensures transparency in identifying AI-assisted development while acknowledging that instances of non-disclosed ChatGPT usage remain outside the scope of this study.

\subsection{Concrete Example}
\begin{figure}[ht]
    \centering
    \includegraphics[width=\linewidth]{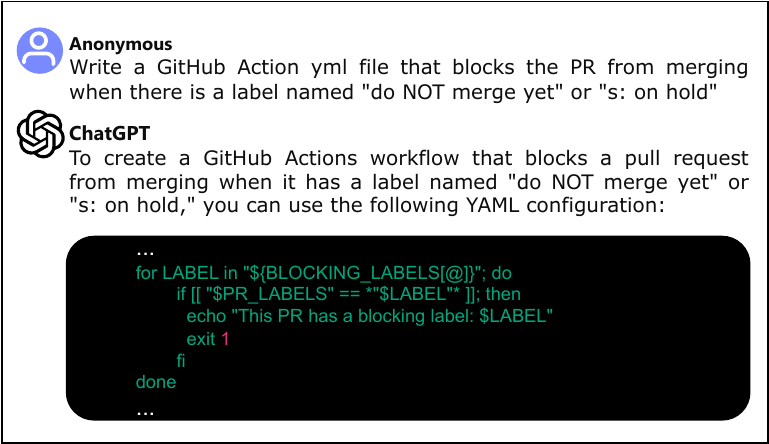}
    \caption{Conversation between developer and ChatGPT. 
    }
    \label{fig:example}
\end{figure}

To ground the terminology used in this study, we present a real-world example of a ``\PA'' scenario. Concrete examples of ``\PN'' and ``\NE'' cases are discussed later in Section~\ref{sec:results-RQ2-RQ3}. In this example, a developer consulted ChatGPT for assistance with a continuous integration issue, and ChatGPT provided a code snippet to address the problem. Figure~\ref{fig:example} shows a portion of this interaction. The full dialogue is available online~\cite{chatpr2405} and is referenced in pull request \#2405~\cite{pr2405} from the \texttt{faker-js/faker} repository.

Analysis of the corresponding diff (Commit: \texttt{62eb120}, Hunk: \texttt{@@ -0,0 +1,31 @@}) shows that the developer integrated the ChatGPT-recommended code verbatim before merging the pull request on September 20, 2023. This case exemplifies direct patch application and is categorized as \texttt{PA}. This example serves to concretely ground the classification used throughout the study and motivates the research questions that examine how frequently such patterns occur and how developers integrate, adapt, or discard AI-generated suggestions in pull request workflows.

Beyond the scenarios where the pull request is merged, we also analyze cases where a pull request is closed without merging, despite developers sharing ChatGPT conversation links within the pull request discussions. In these instances, the developer creates a new branch, modifies files, and submits a pull request following the standard workflow. ChatGPT may have assisted with code suggestions, debugging, or conceptual guidance, yet the pull request is ultimately closed due to factors such as scope misalignment, quality concerns, or project constraints. We refer to this scenario as the Closed Case (\CL). This study investigates the reasons behind \CL, examining how ChatGPT’s involvement influenced closure decisions and whether AI-assisted contributions shaped the final outcome.

\section{Study Design}\label{sec:studydesign}
\begin{figure*}
    \centering
    \includegraphics[width=\linewidth]{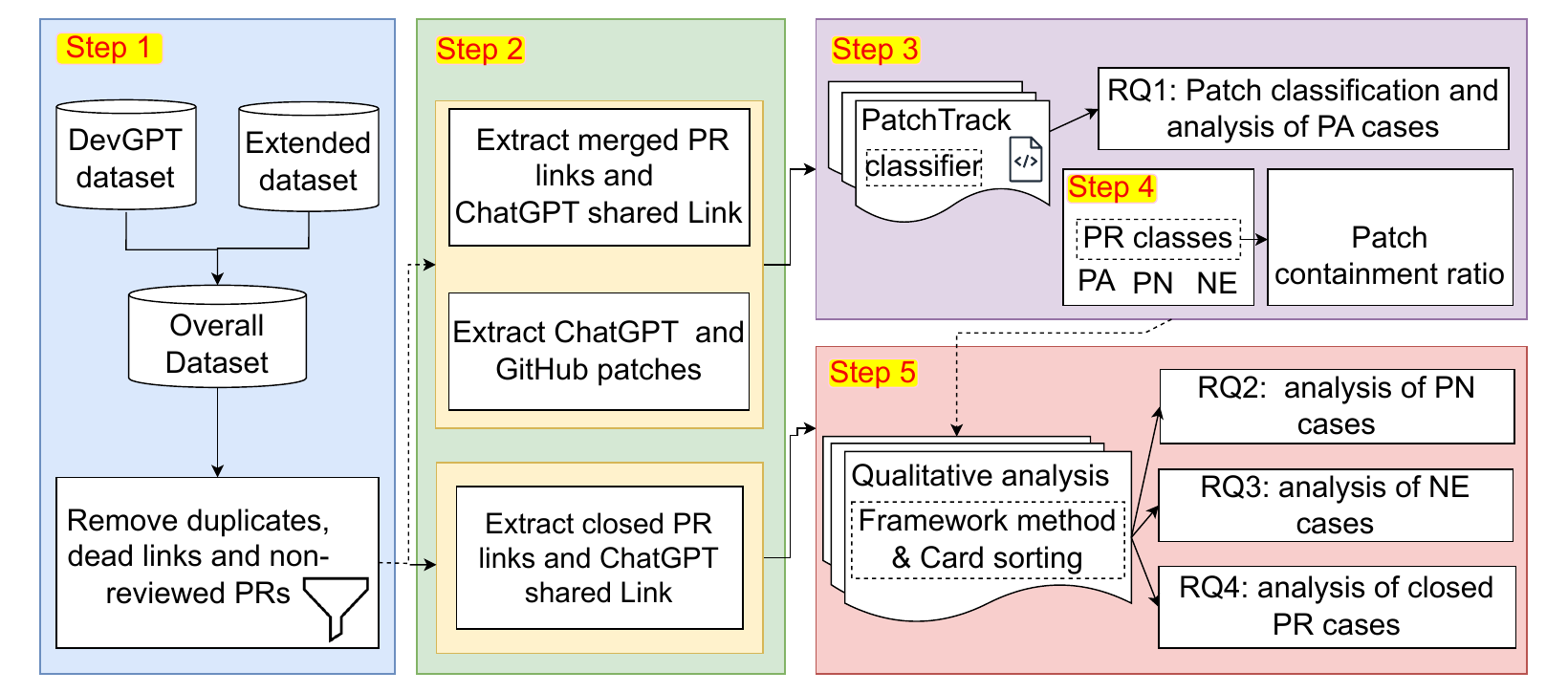}
    \caption{Overview of the study method.} 
    \label{fig:method}
\end{figure*}

This section presents the research method from data collection, tool design, and quantitative and qualitative analysis. The overarching study design is illustrated in Figure~\ref{fig:method}.

\subsection{Research Questions}
\begin{itemize}[leftmargin=*]
\setlength{\itemsep}{0pt} 
    \item \textbf{RQ1:} \textit{How are ChatGPT-generated patches used and integrated in pull requests under self-admitted ChatGPT usage, as observed through patch-level classification?}
    This RQ investigates both the distribution and nature of ChatGPT-generated patch integration in merged pull requests. We use \texttt{PatchTrack} to classify pull requests into three categories: (i) patches that were applied (\PA), (ii) patches that were suggested but not applied (\PN), and (iii) cases where ChatGPT did not generate patches (\NE). In addition to this quantitative view, we conduct a thematic analysis of \PA cases to understand how developers adopt, reshape, or refine ChatGPT’s contributions. Together, these perspectives provide a comprehensive view of AI-generated patch adoption and its role in collaborative development.

    \item \textbf{RQ2:} \textit{Why are ChatGPT-suggested patches in conversations not directly integrated (\PN) into the pull request, and how do developers use ChatGPT’s suggestions in their workflow under the self-admitted ChatGPT usage?}
 Understanding why developers choose not to integrate ChatGPT-generated patches can provide insights into the challenges and constraints that influence AI-assisted development. Furthermore, examining how developers engage with ChatGPT suggestions can shed light on the broader role of AI in software engineering workflows.

\item \textbf{RQ3:} \textit{Under what circumstances does ChatGPT not generate patches in developer interactions (\NE), and how do developers use its responses under the self-admitted ChatGPT usage?} 
    This RQ examines cases where ChatGPT does not suggest direct code patches, exploring the reasons behind such interactions. In addition, it investigates how developers incorporate the responses of ChatGPT - whether for conceptual understanding or other non-patch-related tasks - into their development workflows.

    \item \textbf{RQ4:} \textit{How does ChatGPT influence closed pull requests, and what factors contribute to these outcomes under the self-admitted ChatGPT usage?} 
    This RQ examines the relationship between ChatGPT-assisted contributions and closed pull requests. It aims to explore the factors that lead to pull request closures in cases where ChatGPT-generated content is referenced, providing insights into the broader role of AI-assisted development in collaborative software engineering.

\end{itemize}

\subsubsection{Scope and Unmodeled Factors.}
Our study does not control for prompt formulation or developer expertise, though both can influence the quality and integration of AI-generated code. Prior work shows that incomplete or ambiguous prompts can yield suboptimal suggestions~\cite{ehsani2025promptgaps}, and that developers often adapt or reject AI-generated code based on perceived quality~\cite{grewal2024analyzing, siddiq2024quality}. We instead focus on self-admitted ChatGPT usage in real-world pull requests, enabling grounded analysis of integration behavior without assumptions about prompt content. These unmodeled factors affect all classes in our study (\PA, \PN, \NE, \CL) and are unlikely to bias classification or thematic coding disproportionately. We highlight prompt quality and user expertise as valuable future directions for modeling variation and supporting cross-tool comparisons in AI-assisted development.

\subsubsection{Rationale for Categorization}
We classify ChatGPT-assisted pull requests into four categories: Patch Applied (\PA), Patch Not Applied (\PN), No Patch Generated (\NE), and Closed (\CL). This categorization is central to our study as it enables a structured understanding of how developers engage with AI-generated suggestions. \PA cases allow us to trace direct integration of ChatGPT-generated code. \PN cases highlight friction points such as lack of trust, scope misalignment, or manual refinement that prevent full adoption. \NE cases reveal how developers use ChatGPT for conceptual understanding or planning, even when no code is generated. Finally, \CL cases expose collaborative or technical breakdowns where AI-supported contributions did not result in a merged PR. By segmenting our dataset in this way, we surface distinct usage patterns, support comparative analysis across outcomes, and frame actionable insights for improving AI-assisted pull request workflows.

\subsubsection{Rationale for Descriptive Analysis}
Our quantitative analysis is primarily descriptive due to the heterogeneous and context-dependent nature of AI-assisted pull request interactions. Prior pull request studies have applied inferential models using stable outcome variables such as merge status, review latency, or patch size~\cite{Gousios:ICSE:2014, Tsay-2014, Gousios:ICSE:2016}. 
In contrast, categories such as \PA, \PN, and \NE reflect qualitatively different forms of engagement, including partial reuse, conceptual guidance, and rejection after deliberation, which are not directly comparable across pull requests. These categories represent qualitatively distinct engagement outcomes that cannot be meaningfully modeled using uniform inferential assumptions. We therefore use descriptive statistics to characterize integration patterns and complement them with qualitative analysis to interpret developer decision-making.
Accordingly, our study is exploratory in nature, aiming to characterize integration patterns and developer decision-making rather than to test causal hypotheses.

\subsection{Dataset Construction and Processing}
\label{sec:data-collection}
Figure~\ref{fig:method} provides a visual representation of the dataset extraction and processing workflow. The following steps outline the methodology used to identify, extract, and classify ChatGPT interactions within pull requests, ensuring a systematic and reproducible approach to data collection.

\subsubsection{\textbf{Step 1:} Identification of merged pull requests containing ChatGPT conversation shared links:}

We build on the original DevGPT dataset~\cite{xiao_2023_8304091}. This dataset contains nine snapshots collected from GitHub and Hacker News between July 27, 2023, and October 12, 2023, featuring commits, issues, discussions, pull requests, and threads where developers shared links to ChatGPT conversations. Our study focuses only on the merged and closed (unmerged) pull requests included in these snapshots. The initial DevGPT dataset comprises 98 merged pull requests and 39 unmerged (closed) pull requests.

\textbf{Expanding the DevGPT Dataset}. To ensure a comprehensive and up-to-date analysis, we aimed to enlarge the dataset by including more recent instances of merged pull requests and those that are closed without merging but containing ChatGPT conversation links. Using GitHub API v3~\cite{GitHubAPIv3}, we replicated the original data collection methodology with modifications to accommodate data from October 13, 2023, to August 24, 2024. Our search parameters targeted pull requests that specifically included the keyword ``https:\//\//chat.openai.com/share''. We used a regular expression (``https:\//\//chat.openai.com\//share\//[a-zA-Z0-9-]{36}'') to precisely extract ChatGPT links from pull request descriptions, commits, or review discussions.

This process identified 187 new merged PRs and 14 closed/unmerged PRs. We excluded only PRs whose ChatGPT share links were inactive (e.g., deleted or inaccessible). We did not filter repositories by popularity or overall activity, but required that each repository contain at least one PR with at least one review comment to ensure meaningful discussion context. The final dataset comprises 338 PRs from 255 unique repositories. Summary repository metrics (median stars = 58, median forks = 31, median watchers = 946) indicate that most projects are active and nontrivial; full distributions are presented in Section~\ref{sec:dataset-characteristics} (Fig.~\ref{fig:project_stat}).

\subsubsection{\textbf{Step 2:} \textit{Extracting pull requests and ChatGPT patches}}
The pull request and ChatGPT patches are central for analysis in \textit{RQ1}. Our dataset consists of links to pull requests, ChatGPT links, and patches (if any) extracted from the conversation. The process of extracting patches from ChatGPT conversations was not completely automated, especially in the extended dataset. We noted that OpenAI's updated terms of service (as of November 14, 2023) prohibit users from automatically extracting data directly from their services \cite{openaitoc}.  For this reason, we did not employ any automation or scraping tools on ChatGPT’s website or API.
Instead, we manually converted each shared conversation link obtained in \textit{Step 1} into a local HTML file and performed any necessary parsing only on these static offline copies local without making HTTP requests or automated interaction with OpenAI’s systems. Pull request patches were extracted using GitHub’s \texttt{/pulls/} endpoint and saved as ``diff'' files. At the end of this step, we had two sets of patches for each pull request: ChatGPT and pull request patches. This data served as input for the patch classification phase.

\subsubsection{\textbf{Step 3:} \textit{Patch classification}}
\label{sec:patch-classification}

To answer RQ1, we needed a reliable way to determine whether and how ChatGPT-suggested patches were integrated into pull requests. The goal of this step is to classify each pull request into one of three integration outcomes based on how the AI-generated patch relates to the developer’s final code: Patch Applied (\PA), Patch Not Applied (\PN), or No Patch Exists (\NE). A pull request is labeled \PA if ChatGPT-suggested code appears in the merged changes, \PN if code was suggested but not incorporated, and \NE if no code patch was suggested in the conversation. For brevity, we present only the core classification logic here and defer implementation details, validation procedures, and robustness analysis to Appendix~A.

To enable this analysis at scale, we developed \texttt{PatchTrack}, an automated tool that detects and compares ChatGPT-generated code snippets with pull request diffs. The tool outputs a patch-level label indicating whether the ChatGPT code was directly applied, discarded, or absent, supporting large-scale empirical analysis of developer behavior under self-admitted ChatGPT usage. Patch classification is performed at the hunk level by comparing tokenized representations of ChatGPT snippets with added lines in pull request diffs.

Based on a manual validation study of 230 pull requests, we selected an n-gram size of $n=1$, which achieved the best balance of accuracy, precision, and recall while capturing short and partial code fragments common in ChatGPT responses. Inter-rater agreement between the two authors yielded a Cohen’s Kappa of 0.85, indicating strong reliability. Additional labels used to capture unsupported file types and processing failures are handled internally by the tool and described in Appendix~A, along with detailed descriptions of PatchTrack’s design, normalization procedures, n-gram sensitivity analysis, performance metrics, error sources, and robustness checks.

\subsubsection{\textbf{Step 4:} \textit{Pull request classification}}
The classification of pull requests in \textit{Step 3} follows a straightforward approach. For each class, we aggregate its occurrences within a pull request. If the count of \PA is greater than zero, the entire pull request is classified as \PA. If there are no instances of \PA but there are instances of \PN, the pull request is classified as \PN. If no ChatGPT code snippets were identified in \textit{Step 3}, both \PA and \PN counts for this pull request will be zero, leading to an overall classification of \NE. Pull requests that were closed without merging are classified as \CL, regardless of whether they contain ChatGPT-suggested patches, allowing us to analyze closure-specific factors in RQ4. The analysis provided by \texttt{PatchTrack} shows how the 645 ChatGPT-generated code snippets were compared with 3,486 pull request patches, resulting in precise classifications and valuable insights into the interaction between developers and AI-generated patches. While \texttt{PatchTrack} classifies each individual hunk, our analysis aggregates the results at the pull request level. This is because patch usage decisions (application, revision, rejection, or closure) are typically discussed and resolved within the PR context rather than at the level of individual snippets.

\subsubsection{\textbf{Step 5:} \textit{Qualitative Analysis}}
To address RQ1–RQ4, we conducted a qualitative analysis of pull requests pre-processed and classified by \texttt{PatchTrack}. For RQs, we analyzed cases labeled \PA, \PN, \NE, and \CL, which represent scenarios where ChatGPT suggestions were adopted,  modified, conceptually reused, or ultimately rejected.
For each case, we examined the linked ChatGPT conversation, the surrounding GitHub discussion (up to three developer messages before and after the shared link), and relevant code diffs. This interaction window was selected based on pilot testing, which showed it typically captured sufficient context without including unrelated discussion. Code changes were also reviewed to assess alignment with AI suggestions.  

We applied the Framework Method~\cite{Gale2013}, proceeding through structured summarization, codebook development, and theme refinement. One author and a research assistant independently reviewed overlapping subsets of PR cases and generated structured summaries using a consistent template. Each summary captured the developer’s problem, the ChatGPT recommendation, the developer’s response, the resulting code outcome, and any reviewer feedback. To improve consistency and reduce fatigue for longer threads, we tested GPT-4-generated summary drafts. These were always reviewed and revised by a human analyst. In cases where the model introduced hallucinations or omitted relevant context, the human-authored version was prioritized. All summaries were then reviewed and finalized in discussion with a senior researcher to ensure agreement and interpretive alignment across cases.

All summaries were collaboratively coded and grouped into higher-level themes through card sorting, continuing the three-person team from the summarization phase (one author, a research assistant, and a senior researcher). Initial coding and grouping were performed by the first author and assistant, with discrepancies discussed and resolved in consensus meetings involving the second author and senior researcher. These discussions helped clarify the relationships between surface behaviors and underlying motivations, ensuring interpretive consistency. While we did not compute Cohen’s Kappa due to the qualitative nature of the task, agreement was verified through iterative discussion and convergence on shared meanings. The final thematic structure was applied across the full dataset.\footnote{The research assistant contributed only to manual analysis and theme sorting. They were not involved in study design, interpretation, or writing and therefore do not meet authorship criteria. This follows ACM’s authorship policy: \url{https://www.acm.org/publications/policies/new-acm-policy-on-authorship}}

While our coding was primarily inductive, we consulted prior studies on pull request rejection and review dynamics~\cite{soares2015acceptance,Gousios:ICSE:2014, Gousios:ICSE:2016} to inform our general understanding. These provided background perspective but did not directly guide our codes, as they do not account for the unique characteristics of AI-assisted contributions. Because AI-generated patches introduce distinct risks and social dynamics, such as hallucinated output, prompt ambiguity, developer over-reliance on AI, and uncertainty in trust attribution, we extended our codebook by open coding 20 representative cases to surface patterns grounded in the data.
Examples of structured summaries and the code-to-theme mapping are included in our replication package~\cite{patchtrack:2025} to support transparency and reproducibility.

\section{Results and Discussion}
\label{sec:results}

This section presents the key findings of our study. We begin by characterizing the repositories and pull requests that form the basis of our dataset. This provides essential context for interpreting how ChatGPT-generated patches are used in real-world development workflows. We then answer each of our research questions in turn, starting with RQ1, which quantifies the integration of ChatGPT-generated code across pull requests.

\subsection{Repository Characteristics}
\label{sec:dataset-characteristics}
\begin{figure}[!h]
    \centering
    \includegraphics[width=\linewidth]{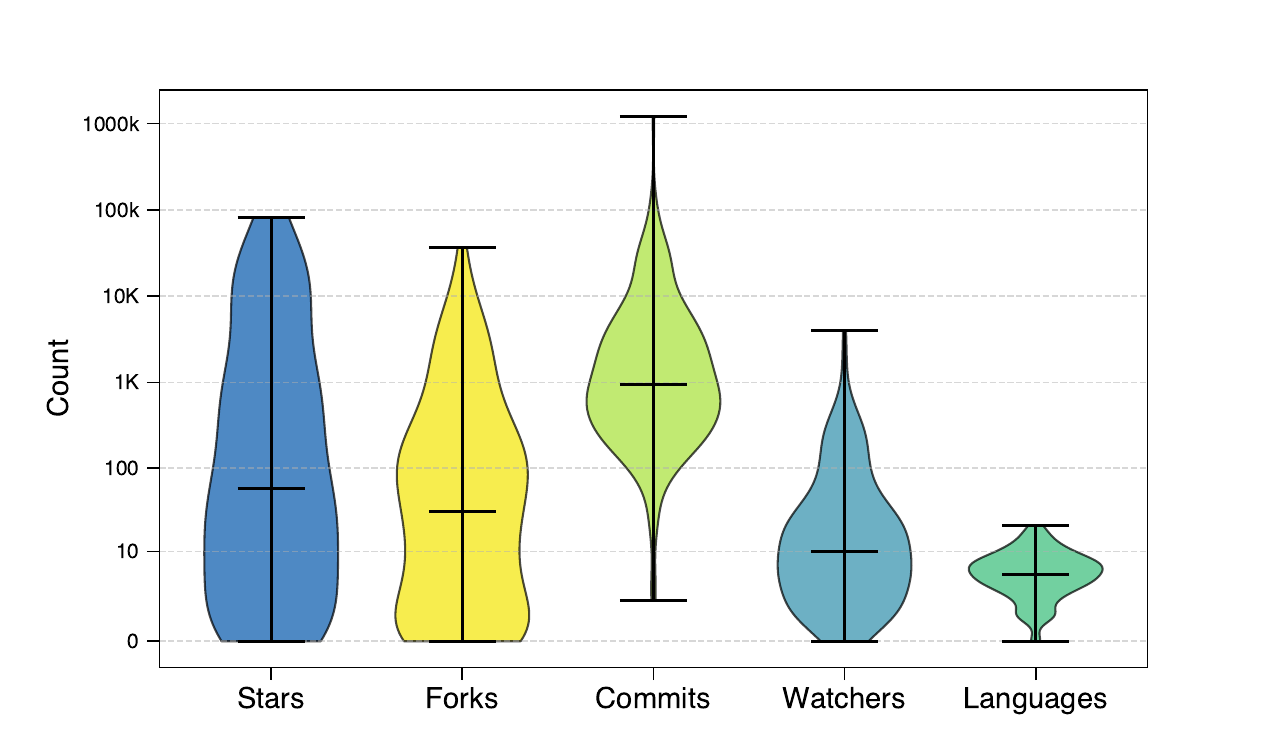}
    \caption{Characteristics of the repositories curated in the dataset.}
    \label{fig:project_stat}
\end{figure}

The 338 pull requests in our dataset come from 255 unique GitHub repositories that vary widely in maturity, popularity, and technical focus. Figure~\ref{fig:project_stat} summarizes core repository attributes. Popularity indicators are long-tailed: stars range from 0 to $\sim$81{,}000 (median: 58) and forks from 0 to 36{,}655 (median: 31). Commit histories are likewise highly skewed, with counts peaking above 1.2\,million (median: 946). The language panel further indicates coverage across diverse implementation ecosystems. These attributes indicate that our dataset includes both niche and high-profile projects. Prior studies have used similar repository-level metrics to capture ecosystem diversity and developer participation~\cite{Weeraddana2023,10298470,businge:saner:2022}.

In terms of language diversity, the dataset covers 123 distinct programming languages. The most frequently observed languages include Shell (85.5\%), JavaScript (82.0\%), HTML (62.0\%), TypeScript (50.0\%), Python (54.1\%), and CSS (54.0\%), reflecting a strong representation of web development and scripting tasks. Infrastructure and configuration files are also well represented, such as Dockerfile (49.0\%) and Makefile (33.0\%). Compiled and system-level languages such as Java (19.0\%), C (18.4\%), and C++ (14.5\%) are present but less dominant.

While this linguistic variety highlights the breadth of our dataset, its distribution diverges from GitHub’s overall language trends. According to GitHub’s 2024 Octoverse report~\cite{github_octoverse_2024}, Python has surpassed JavaScript as the most used language, with Java, C, and C++ also ranking among the top. In contrast, our dataset overrepresents Shell, JavaScript, and other front-end and scripting technologies. This suggests that developers are more likely to use ChatGPT in tasks that involve scripting, web development, and configuration management, domains that favor automation and well-defined coding patterns. By contrast, areas that demand deeper domain knowledge or complex architectural decisions (e.g., backend systems, data science) may be less conducive to AI-assisted contributions.

These findings suggest that current ChatGPT usage in pull request workflows is concentrated in specific development contexts. Broader conclusions about generative AI adoption will require datasets with greater alignment to GitHub’s full language ecosystem. Future work could address this by stratifying repositories according to GitHub language distributions and exploring AI-supported development practices across a wider range of software domains.

\subsection{RQ1: Classification of Patch Use in Pull Requests}\label{sec:result-rq1}
\noindent \textit{RQ1: How are ChatGPT-generated patches used and integrated in pull requests under self-admitted ChatGPT usage, as observed through patch-level classification?}
\\
This research question investigates the distribution of ChatGPT-assisted pull requests across three integration categories: Patch Applied (\PA), Patch Not Applied (\PN), and No Patch Exists (\NE). Using the final configuration of \texttt{PatchTrack} (\(n=1\)), we analyzed 285 pull requests, resulting in the following classification:
\vspace{-0.5em}  

\begin{figure}[!ht]
    \centering
    \includegraphics[width=1\linewidth]{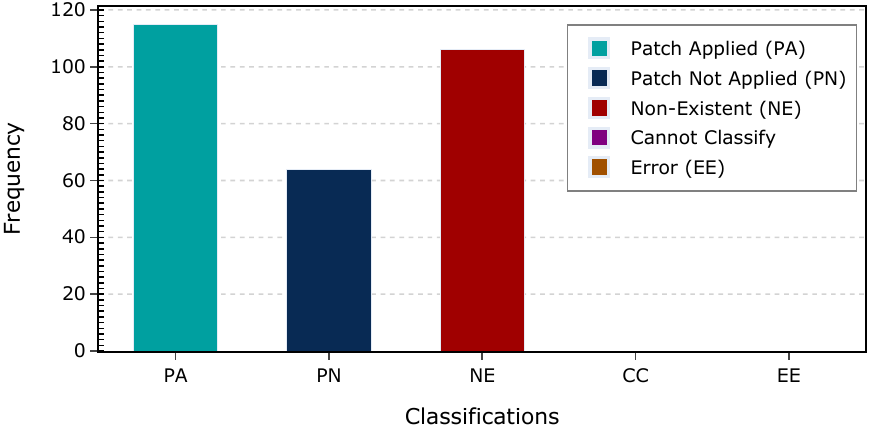}
    \caption{Distribution of Patch Classification Results}
    \label{fig:classification}
\end{figure}
\vspace{-1.5pt}  

\begin{itemize}
\item \textbf{Patch Applied (\PA)}: 116 cases (40.7\%)
\item \textbf{Patch Not Applied (\PN)}: 63 cases (22.1\%)
\item \textbf{No Patch Exists (\NE)}: 106 cases (37.2\%)
\end{itemize}

Figure~\ref{fig:classification} visualizes this distribution. While many developers adopted ChatGPT-generated patches, others rejected or heavily modified them, and in some cases, ChatGPT did not generate a patch at all. These outcomes reflect varying levels of trust, relevance, and integration effort, which we explore further in RQ2–RQ4.

To assess the sensitivity of these results to potential PatchTrack misclassification, we recomputed the main RQ1 statistics using only the manually validated subset of 230 pull requests. 
The relative distribution and ordering of the \PA, \PN, and \NE classifications remained consistent with those observed in the full dataset, and no classification shifted from minority to majority. The median integration rate for \PA cases also remained unchanged. This indicates that the aggregate patterns reported for RQ1 are robust to possible classification noise.

\subsubsection{Patch Classification}
We analyzed 645 ChatGPT-generated code snippets and 3,486 patches across 285 pull requests. Each snippet was compared against all patches in its corresponding PR using \tool, which employs token-level matching to determine integration. This process enabled classification into \PA, \PN, or \NE, as described in Section~\ref{sec:patch-classification}.

As validated in our methodology (Section~\ref{sec:patch-classification}), using an n-gram size of \(n=1\) yielded the highest accuracy, precision, recall, and F1 score, due to its sensitivity to short code fragments. This configuration was used for all reported results. Figure~\ref{fig:classification} summarizes the resulting distribution of PRs across the three integration categories.

\paragraph{Percentage of ChatGPT Patches Applied}  

To assess how much of ChatGPT’s suggested code is integrated into pull requests, we compute the percentage of ChatGPT patches applied. This metric helps evaluate the degree to which developers adopt AI-generated suggestions when resolving issues or adding features. By linking overlap percentage to classification labels, we can systematically evaluate developer adoption behavior and determine how self-admitted ChatGPT usage translates into actual code integration. For example, a high ratio indicates direct reuse of ChatGPT’s output, while a low ratio may suggest heavy modification or rejection. This  directly supports our goal of classifying integration outcomes. We use Jaccard’s containment ratio~\ref{eq:jaccard} to quantify the overlap between tokens in the ChatGPT-suggested patch ($t_x$) and the final integrated code ($t_y$):

\begin{equation}\label{eq:jaccard}
Jacc(t_x, t_y) = \frac{|t_x \cap t_y|}{|t_x|} \times 100%
\end{equation}

\begin{figure}[!ht]
    \centering
    \includegraphics[width=1\linewidth]{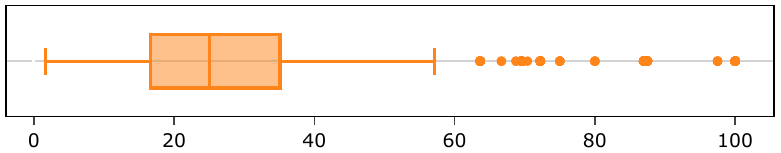}
    \caption{Boxplot showing the percentage of lines of code from the ChatGPT patches that are integrated into pull requests.}
    \label{fig:boxplot}
\end{figure}

Here, $t_x$ and $t_y$ are token streams from ChatGPT and GitHub patches, respectively. This formulation, adopted from the \texttt{Redebug} study~\cite{6234404}, measures the proportion of ChatGPT’s code that appears in the final merged pull request. Figure \ref{fig:boxplot} illustrates the distribution of integration percentages for the classification \PA. The median integration rate is 25\%, indicating that, on average, a quarter of the ChatGPT-suggested code is adopted in merged pull requests.

Figure~\ref{fig:boxplot} reports the proportion of ChatGPT-suggested code that overlaps with the final pull request. This metric captures the degree of reuse between AI-generated and integrated code, but does not distinguish whether overlapping lines correspond to newly added code or modifications of existing code. As such, the figure should be interpreted as indicating reuse intensity rather than the semantic nature of the changes.

\noindent Developers integrate ChatGPT patches with varying degrees of modification, assessing their relevance based on project needs. While some adopt AI-generated patches wholesale, others incorporate only selective portions. This variation reflects how developers balance AI-assisted suggestions with project-specific constraints such as maintainability, readability, and adherence to coding standards. The distribution of integration percentages, as shown in Figure~\ref{fig:boxplot}, highlights this diversity in adoption, with some pull requests fully incorporating ChatGPT-generated code, while others retain only fragments. The presence of outliers in the distribution underscores the exception rather than the norm: while a few developers fully adopt ChatGPT-generated patches, the majority incorporate only parts of the suggestions, often modifying them extensively to fit project-specific requirements.

The median integration rate of 25\% (Q2) suggests that full adoption of ChatGPT-generated patches is uncommon, with most pull requests retaining only a fraction of the AI-suggested code. The third quartile (Q3) at 38\% further reinforces that even in cases with higher integration, significant modifications are often required. These findings indicate that developers frequently treat the ChatGPT output as a \textbf{starting point rather than a final implementation}, selectively incorporating code that aligns with project needs. 

To better understand this variation, we selected four pull requests representing different levels of integration, with one pull request chosen from each quartile:  
(i) \textit{low (0–25\%]}, where only minor elements of the suggested patch were retained, often requiring substantial modifications or alternative implementations;  
(ii) \textit{lower-median (25–50\%]}, where developers incorporated portions of the patch but refined it significantly;  
(iii) \textit{upper-median (50–75\%]}, where a larger portion of the patch was adopted with some modifications; and  
(iv) \textit{high (75–100\%]}, where the AI-generated patch was largely accepted with minimal changes.  
Each quartile follows the standard convention where the lower bound is excluded and the upper bound is included. These cases were selected through stratified random sampling to ensure balanced representation across integration levels. By analyzing these strata, we illustrate real-world scenarios of ChatGPT-assisted patch integration, shedding light on why some patches are fully applied while others are selectively merged or heavily revised. These examples are illustrative and not intended for theory-building, so theoretical saturation was not a goal.

\subsubsection{Qualitative Analysis -- RQ1}
\label{sec:results-RQ1-qual}
\noindent To complement the quantitative analysis of patch adoption rates, we conducted a thematic analysis of the validated \PA cases. Although 90 cases were initially sampled, four were later reclassified as \PN and three from \PN to \PA, resulting in a final set of 89 \PA cases (see Section~\ref{sec:patch-classification} for details on misclassifications). Coding followed the structured framework described in our methodology. Agreement was reached through coder discussion and convergence on six themes.

\noindent\textbf{RQ1 qualitative themes.}
To support interpretation of the theme distributions reported below, we briefly summarize the six qualitative themes used to code ChatGPT’s contribution in pull requests.
These themes capture distinct modes of developer engagement with AI output, ranging from direct code reuse (\textit{Direct Implementation}), partial or selective reuse (\textit{Selective Extraction}, \textit{Structural Integration}), and iterative modification (\textit{Iterative Refinement}), to constraint-driven adaptation (\textit{Constraint Handling}) and non-code influence such as conceptual guidance or documentation support (\textit{Knowledge Support}).
Full operational definitions and coding criteria are provided in Appendix~B.

Table~\ref{tab:theme-summary} summarizes how themes distribute across integration ranges. For brevity, we present one representative example per quartile below. Each instance is labeled \texttt{PA-X} (where \texttt{X} is the instance number). Full case summaries, coding assignments, pull request links, and ChatGPT–developer conversation links are available in the replication package~\cite{patchtrack:2025}.

\begin{table}[h]
\centering
\caption{Summary of themes across integration ranges.}
\begin{tabular}{lccccc}
\toprule
\textbf{Theme} & \textbf{(0--25\%]} & \textbf{(25--50\%]} & \textbf{(50--75\%]} & \textbf{(75--100\%]} & \textbf{Total} \\
\midrule
Constraint Handling     & 4  & 3  & 3  & 0 & 10 \\
Direct Implementation   & 0  & 0  & 0  & 3 & 3  \\
Iterative Refinement    & 10 & 3  & 7  & 6 & 26 \\
Knowledge Support       & 4  & 1  & 5  & 3 & 12 \\
Selective Extraction    & 8  & 9  & 0  & 1 & 18 \\
Structural Integration  & 8  & 9 & 1  & 1 & 19 \\ \midrule
\textbf{Total}             & \textbf{34} & \textbf{25} & \textbf{16} & \textbf{14} & \textbf{89} \\
\bottomrule
\end{tabular}
\label{tab:theme-summary}
\end{table}

\noindent \faCheckSquareO \textbf{Low Integration (0–25\%):} 
This quartile (34 cases) is characterized by \textit{Structural Integration} (8 cases) and \textit{Selective Extraction} (8 cases) as the most distinctive behaviors for minimal carryover of AI code, even though \textit{Iterative Refinement} appears slightly more often numerically (10 cases). We emphasize \textit{Structural Integration} and \textit{Selective Extraction} here because they most clearly explain \textit{why} integration remained low: developers either reshaped ChatGPT’s output to fit project frameworks or extracted only fragments of functionality, discarding the rest. These patterns reflect how AI output typically required substantial reshaping before it could be meaningfully integrated in this range (counts from Table~\ref{tab:theme-summary}).
Example 1 (PA-27)--Case ID: PA-27; Theme: \textit{Structural Integration} illustrates this pattern: 
In pull request \#185~\cite{pr185} of the \texttt{pokt-network/poktroll} repository, the developer asked ChatGPT\footnote{\href{https://chat.openai.com/share/d68981db-26e1-431c-841d-2bb31096c0c9}{Example 1 link to ChatGPT conversation} -- Accessed: 2025-02-17} if \texttt{Go} had a built-in function to remove leading spaces from multiline strings while maintaining code readability. ChatGPT suggested an approach using \texttt{strings.Split}, \texttt{strings.TrimLeft}, and \texttt{strings.Join} to process each line and remove leading spaces. The developer then requested a refined solution that encapsulated this logic into a reusable helper function, \texttt{trimString}, to improve readability. 
ChatGPT provided a full \texttt{Go} program, but only its helper function was integrated. Of the suggested 31 LOC, just 7 LOC (22.6\%) were adopted. The minimal adoption reflects the developer’s selective integration of only the core functionality needed, as the existing project structure did not require a complete rewrite.

\vspace{5pt}
\noindent \faCheckSquareO \textbf{Lower-median Integration (25--50\%):}  
This quartile (25 cases) is dominated by \textit{Selective Extraction} (9 cases) and \textit{Structural Integration} (9 cases), with \textit{Constraint Handling} and \textit{Iterative Refinement} also present but less frequent (counts from Table~\ref{tab:theme-summary}). Developers in this range often incorporated substantial portions of ChatGPT’s suggestions, but not wholesale: they selectively retained useful fragments or adapted the AI output to align with coding conventions. Integration percentages therefore remained below 50\%, reflecting the tendency to treat ChatGPT’s patch as scaffolding rather than final code.  
Example PA-36; Theme: \textit{Selective Extraction} illustrates this pattern:  
In pull request \#7123~\cite{pr7123} of the \texttt{Mudlet/Mudlet} repository, a crash occurred when double-clicking a word in \texttt{TTextEdit}, traced to an \textit{out-of-bounds access} in text-selection logic. After reviewing the \textit{stack trace}, the developer consulted ChatGPT\footnote{\href{https://chat.openai.com/share/cf97d3d8-e419-4878-9a2b-87409195340a}{Example 2 link to ChatGPT conversation} -- Accessed: 2025-02-17}, which proposed a structured patch: (i) add boundary checks for \texttt{yind}, (ii) break loops when indices become invalid, and (iii) refactor selection logic for robustness. The 39-LOC patch addressed the crash, but only 18 LOC (46.2\%) were integrated. A reviewer suggested simplifying the control flow by \textit{negating conditions to reduce indentation}, prompting further edits. The adopted code retained the essential boundary checks while discarding or refactoring the rest, showing how developers extracted only the useful fragments of ChatGPT’s logic while reshaping them to improve readability and project alignment.

\vspace{5pt}
\noindent \faCheckSquareO \textbf{Upper-median Integration (50--75\%):}  
This quartile (16 cases) is characterized by \textit{Iterative Refinement} (7 cases) as the most common theme, followed by \textit{Constraint Handling} (3 cases) and \textit{Knowledge Support} (5 cases) (counts from Table~\ref{tab:theme-summary}). Developers in this range integrated a majority of ChatGPT’s suggestions, but usually with targeted refinements rather than direct adoption. Reviewer input often drove these refinements, such as clarifying class names, reorganizing logic, or improving readability. Integration percentages in this quartile therefore reflect substantial reuse of ChatGPT code, tempered by human oversight to align with project conventions.  
Example PA-73; Theme: \textit{Iterative Refinement} illustrates this pattern:  
Pull request \#279~\cite{pr279} in the \texttt{nylas/nylas-python} repository introduced support for a new ``free-busy'' API endpoint in the Nylas SDK, enabling users to check calendar availability for specific email addresses. A key challenge was designing a suitable data model to represent the API response, which could include both availability data and error messages. To address this, the developer consulted ChatGPT\footnote{\href{https://chat.openai.com/share/51f3aa63-d8aa-4ff7-aca1-608fcf9ab9ee}{Example 3 link to ChatGPT conversation} -- Accessed: 2025-02-17} for guidance on structuring Python dataclasses. ChatGPT suggested: (i) defining separate data models for distinct response components, (ii) using Python’s \texttt{dataclass} module for maintainability, and (iii) employing \texttt{Union} types to handle multiple object types within a list. The proposed patch contained 25 LOC, of which 15 LOC (60\%) were integrated. The main adjustment involved renaming the \texttt{error} class to \texttt{freeBusyError} for clarity, following reviewer feedback. This case shows how AI-generated code can align closely with project needs while still requiring human-led refinements to enhance readability and maintainability.  
 
\vspace{5pt}
\noindent \faCheckSquareO \textbf{High Integration (75--100\%):}  
This quartile (14 cases) is dominated by \textit{Direct Implementation} (3 cases) alongside instances of \textit{Iterative Refinement} (6 cases) and \textit{Knowledge Support} (3 cases) (counts from Table~\ref{tab:theme-summary}). Developers in this range adopted the ChatGPT output with little modification, often because AI-generated code already aligned with project requirements. When changes occurred, they were minor and stylistic rather than structural. These cases highlight scenarios in which ChatGPT produced production-ready code that required minimal oversight before integration.  
Example PA-82; Theme: \textit{Direct Implementation} illustrates this pattern:  
Pull request \#2059~\cite{pr2059} of the \texttt{alshedivat/al-folio} repository extended a Jekyll plugin to support fetching posts from a manually specified list of URLs in addition to an RSS feed. Initially, the plugin retrieved posts only via RSS, but the developer aimed to enhance functionality by incorporating direct URL input. To achieve this, they consulted ChatGPT\footnote{\href{https://chat.openai.com/share/24432d24-36a7-4d6f-a5c0-d7e5142f68cd}{Example 4 link to ChatGPT conversation} -- Accessed: 2025-02-17}, focusing on three key aspects: (i) metadata extraction, (ii) correct handling of published dates, and (iii) improved logging readability. ChatGPT refactored the Ruby plugin to introduce modular functions, ensured robust date handling with \texttt{Time} objects, improved log clarity through \texttt{puts}, and extracted date parsing into a dedicated \texttt{parse\_published\_date} function. The patch consisted of 19 LOC, all of which were integrated as-is (100\%), since the output aligned perfectly with project needs. The developer’s enthusiasm, expressed at the end of the interaction (\textit{\ul{``Awesome! I have a feeling we’ll write a lot of code together from now on ;)''}}), underscores how ChatGPT’s response was fully applicable without requiring further adjustments.  

\subsubsection{Discussion and Implications -- RQ1}\label{sec:discussion-rq1}

Our findings show that while ChatGPT-generated patches provide value, full adoption remains uncommon, with a median integration rate of 25\%. Developers most often treated ChatGPT’s output as a \textit{starting point} rather than a final implementation, selectively integrating elements that aligned with project-specific needs. Thematic analysis of 89 validated \PA cases revealed four dominant integration patterns across quartiles: \textit{Selective Extraction}, \textit{Structural Integration}, \textit{Iterative Refinement}, and \textit{Direct Implementation}. These patterns explain not only how much of ChatGPT’s code was reused but also why integration remained partial or complete.

In low-integration cases (0--25\%), developers retained only minor portions of ChatGPT’s suggestions, typically extracting fragments or reshaping code to fit project frameworks. Moderate integration cases (25--75\%) revealed substantial reuse, but usually with targeted refinements, such as simplifying control flow, renaming classes, or aligning with architectural constraints, often informed by reviewer feedback. High-integration cases (75--100\%) occurred when ChatGPT’s output already conformed to project conventions, enabling near-direct adoption with minimal oversight. These variations highlight that ChatGPT accelerates exploration but rarely produces production-ready patches without human adaptation.

The predominance of low- and moderate-integration cases underscores ongoing challenges in AI-assisted development, including misalignment with project coding styles, maintainability concerns, and trust calibration. This suggests a need for AI tools that incorporate greater contextual awareness, generating code that requires less reshaping. At the same time, high-integration cases demonstrate AI’s potential to streamline the pull request process when outputs align with project expectations, reducing the effort of implementation.

Importantly, our study focuses on self-admitted ChatGPT usage (SACU), which means that all the cases analyzed involve developers explicitly stating their use of ChatGPT within pull request discussions. This ensures transparency in identifying AI-assisted contributions, but also introduces a limitation: \textit{we do not capture instances where ChatGPT was used privately or without disclosure}. The question of whether undisclosed ChatGPT usage follows a different pattern of patch integration, which could influence the observed distribution of \PA, \PN, and \NE cases remains open.

These findings resonate with the longstanding vision of the ``Programmer’s Apprentice''~\cite{rich1990programmers}, positioning ChatGPT not as an autonomous coder but as a junior collaborator that supports debugging, scaffolding, and refinement while leaving final decisions to developers. By classifying adoption behaviors across granular integration levels, we show that ChatGPT functions primarily as a decision-support tool, shaping but not replacing human judgment in pull request workflows.

Finally, this analysis establishes a foundation for the subsequent research questions. While RQ1 quantifies \textit{how} ChatGPT-generated patches are integrated and characterizes adoption patterns, RQ2–RQ4 explore the complementary cases: why patches are not applied (\PN), why no patches are suggested (\NE), and why some pull requests close despite ChatGPT’s involvement (\CL).

\vspace{5pt}
\noindent\textbf{Connection to Prior Work:}
Our findings complement and extend prior studies on AI-assisted code generation~\cite{huang2024generative, menzies2020software, ozpolat2023ai_tools}, code review~\cite{tufano202:IEEEPress, xiao2024generative}, and pull request workflows~\cite{Gousios:ICSE:2014, Tsay-2014, Gousios:ICSE:2016}. While prior work has quantified productivity benefits or patch acceptance~\cite{Jiang:2023, Guo:2024, Deng:2024}, few studies have examined how AI-generated patches are actually evaluated, reshaped, or rejected within the collaborative dynamics of pull requests. Our contribution builds on recent efforts to capture patch-level decisions~\cite{grewal2024analyzing, siddiq2024quality}, but goes further by stratifying cases across integration quartiles and classifying behavioral themes. In doing so, we show that generative AI functions less as a deterministic patch source and more as a catalyst for developer reasoning and selective adoption.

\begin{custombox}
\faLightbulbO \hspace{0.01in} \textbf{Summary of RQ1 Results}: {\textit{ChatGPT-generated patches are seldom adopted wholesale. The median integration rate is 25\%, with adoption patterns varying across quartiles. Our qualitative analysis of 89 cases revealed six recurring themes, with most developers engaging in \textbf{iterative refinement}, \textbf{structural integration}, or \textbf{selective extraction} rather than direct implementation. These findings show that developers consistently treat AI output as a starting point, adapting it to project-specific constraints and reviewer feedback. \texttt{PatchTrack} enabled this analysis by classifying patch use with 97.5\% accuracy and surfacing distinct adoption patterns. Together, these insights motivate RQ2--RQ4, which explore why patches are rejected, why some PRs lack AI-suggested patches, and why others close without adoption despite ChatGPT involvement.}}
\end{custombox}

\subsection{Results and Discussion -- RQ2, RQ3 \& RQ4}
\label{sec:results-RQ2-RQ3}

\noindent This section builds on the validated dataset from RQ1, where 230 pull requests were manually reviewed to assess patch integration outcomes: 90 \PA, 56 \PN, and 84 \NE. We now examine the \PN and \NE subsets to answer RQ2 and RQ3. For RQ4, we independently sampled 47 closed pull requests from a total of 53 using a 95\% confidence level and a 5\% margin of error.

Throughout this analysis, we do not interpret patch non-adoption (\PN) or pull request closure (\CL) as evidence of low code quality; rather, we treat these outcomes as reflecting contextual constraints, integration effort, and maintainer decision-making during review.

Thematic coding was conducted using a structured framework based on our research questions. While we did not compute Cohen’s Kappa due to the interpretive nature of the task, agreement was reached through coder discussion and convergence on final themes. Details of the coding procedure and reviewer roles are described in Section~\ref{sec:data-collection}, Step 5.

\noindent\textbf{Qualitative themes for RQ2--RQ4.}
Across RQ2, RQ3, and RQ4, we identify recurring themes that characterize how ChatGPT influenced developer decision-making when patches were not applied (\PN), no patches were generated (\NE), or pull requests were closed (\CL). For each research question, we introduce a brief inline characterization of the relevant themes before presenting illustrative cases, focusing on how ChatGPT shaped methodology, design choices, documentation and communication, debugging and optimization, or was constrained by technical, procedural, or scope-related factors. Full operational definitions for all themes are provided in Appendix~C, with detailed coding criteria and additional examples available in the replication package~\cite{patchtrack:2025}.

For brevity, we present one representative instance per theme in the results sections. Additional examples and detailed summaries are available in the replication package \cite{patchtrack:2025}, which includes all summaries, identified themes, pull request links, and corresponding ChatGPT-developer conversation links.
\noindent Each instance is labeled with a unique identifier: PN-X (X being the instance number) for patches not applied, NE-X for cases where no patches were suggested, and CL-X for closed pull requests. These identifiers support cross-referencing with the replication package.

We first present the results, followed by the implications at the end of each research question. In the results, we provide concrete examples and notable excerpts from pull request discussions. Themes are highlighted in \textbf{bold fonts}, conversations are presented in \textit{italics}, keywords related to a particular theme are \underline{underlined}, and longer messages are abbreviated with ellipses [...]. For each theme, we discuss the implications for practitioners and/or researchers using the icons \faLaptop \hspace{0.02in} and \faSearch \hspace{0.02in}, respectively.

\subsection{Results and Discussion -- RQ2} 
\begin{figure}[ht]
    \centering
    \includegraphics[width=\linewidth]{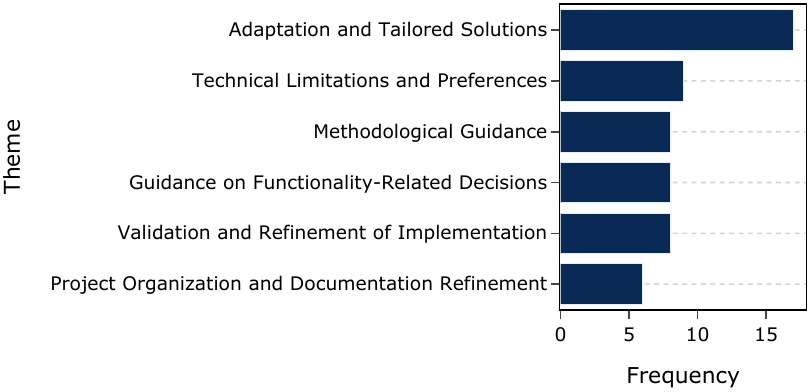}
    \caption{Showing the distribution of themes in the patch not applied (\PN) class.}
    \label{fig:pn-class}
\end{figure}

\noindent \textit{Why are ChatGPT-suggested patches in conversations not directly integrated (\PN) into the pull request, and how do developers use ChatGPT’s suggestions in their workflow under the self-admitted ChatGPT usage?}
In our analysis, \tool initially classified 63 instances as patch not applied (\PN). To validate the accuracy of this classification, we manually reviewed a sample of 55 \PN instances, selected using a 95\% confidence level with a 5\% margin of error, as detailed in Section~\ref{sec:patch-classification}. During this review, we identified 3 false positives that were incorrectly labeled as \PN and 4 false negatives from the patch applied (\PA) set that should have been categorized as \PN. After applying these corrections, we obtained a final set of 56 validated \PN instances for thematic analysis. The analysis of these 56 instances revealed six distinct themes, as illustrated in Figure~\ref{fig:pn-class}.

\noindent \faCheckSquareO \textbf{Adaptation and tailored solutions:} This emerged as the most prevalent theme, with 17 of 56 cases showing that developers adapted the conceptual insights of ChatGPT to fit specific project requirements. For example, in pull request \#12 \cite{pr12} -- PN-19, a developer sought a regular expression from ChatGPT to match the ULID format in the \texttt{laravel-json-api/core} project. The developer initially considered ChatGPT's regex: \textit{``Like I said, I didn't find an 'official' regex for the ULID. ChatGPT gave me this: [...]''}. However, rather than directly applying the AI-generated solution, the developer aligned the implementation with Laravel's native method for handling ULIDs. This decision was influenced by the comment of a reviewer: \textit{\ul{``I think we should match what Laravel does''} for route regexes for ULIDs. Looking at this docs: [...]}. The author acknowledged the feedback: \textit{``[...] good idea [...]. Change made!''} and updated the code to ensure consistency within the Laravel ecosystem. This example highlights how developers use ChatGPT as a starting point, but ultimately refine AI-generated suggestions to conform to established coding conventions and project-specific constraints.

\noindent \faCheckSquareO \textbf{Technical limitations and preferences:} 
This was the second most prominent theme, with 9 of 56 instances demonstrating how ChatGPT’s patches were not applied due to technical constraints or developer preferences for alternative solutions. In these cases, developers faced trade-offs in implementation strategies, either due to limitations in the existing system or subjective choices regarding the best approach to a problem.
A representative example is pull request \#171 \cite{pr171} within the \texttt{ory/elements} project (PN-11), where developers attempted to disable a button when submitting the form. A reviewer sought ChatGPT’s input on a possible fix, asking: \textit{“Wouldn't the fix just be to change click to submit? [...]”} However, another reviewer pointed out a critical issue with this approach: \textit{\ul{“This doesn't actually work. The reason is that the click event fires before the submit event} [...] but IMO we should just get this right [...]”} The pull request author ultimately acknowledged that resolving the issue properly would require additional time and effort, stating: \textit{“[...] just don't want to spend time on getting this right, right now [...]”} Upon further examination, we found that the proposed solution was reverted because disabling the button caused unintended side effects, such as stopping event propagation. Given the minor impact of the issue, the team decided to focus on more pressing updates rather than implementing a workaround.

\noindent \faCheckSquareO \textbf{Methodological Guidance:} 
We identified 8 of the 56 instances in which ChatGPT influenced developers' decision-making strategies by suggesting optimizations or best practices rather than providing direct fixes.
For example, in pull request \#5063~\cite{pr5063} within the \texttt{darklang/dark} project, a developer attempted to resolve an issue related to function composition. A reviewer, finding the initial implementation overly complex, consulted ChatGPT, which suggested a more concise approach. The reviewer commented: \textit{``While this works, it feels like overkill. \ul{I asked ChatGPT about how best to do this, and it suggested using (int)c}. That seems to me like it would work. ChatGPT-link[...]''}. This led the pull request author to reassess the solution, ultimately adopting a simplified version influenced by ChatGPT’s recommendation.

\noindent \faCheckSquareO \textbf{Guidance on Functionality-Related Decisions:} 
Developers often seek to modify or enhance specific functionality in software projects, often consulting ChatGPT for insights on best practices or alternative implementations. This theme captures instances where ChatGPT’s recommendations influenced functionality-related decisions, even when no direct patches were applied. 
From our dataset, we identified 8 of the 56 instances in this category. A representative example is pull request \#522 \cite{pr522} within the \texttt{sveltejs/learn.svelte.dev} repository (PN-13). In this case, the pull request author sought guidance on the correct \texttt{ARIA} role for an interactive \texttt{SVG} acting as a pointer in a \texttt{Svelte} application. The conversation focused on whether to use ``\texttt{role=none}'' or ``\texttt{role=presentation}.'' ChatGPT provided contextual insights on \texttt{ARIA} accessibility, suggesting that ``\texttt{role=presentation}'' would be more appropriate based on the functionality of the \texttt{SVG}s. The pull request description explicitly references ChatGPT’s recommendation: \ul{\textit{``presentation'' might be better suited here according to ChatGPT: [...]}}. 
The integrator acknowledged the contribution with a brief confirmation: \textit{“Thank you!”} and merged the pull request. This case illustrates how ChatGPT's suggestions can shape design decisions and accessibility improvements, with maintainers providing the final validation before integration.
 
\noindent \faCheckSquareO \textbf{Validation and Refinement of Implementation}: 
Developers often use ChatGPT to \textit{validate their reasoning and refine their decision-making process} rather than applying AI-generated patches directly. In 8 of 56 instances, the interactions focused on seeking conceptual clarification or evaluating alternative implementations before finalizing a solution. This indicates that AI tools play a role not only in code generation, but also in \textit{developer cognition and reasoning during pull request workflows}. 
A representative example is pull request \#5058~\cite{pr5058} -- PN-28, where a developer was considering implementing the \texttt{List.filter} function using recursion versus \texttt{List.fold} in a functional programming context. A reviewer commented:
\textit{\ul{``I was thinking about whether recursive implementations were more appropriate here}, since this can also be implemented with List.fold. [...] You might enjoy what ChatGPT told me about it: [...]''} 
This prompted further discussion about theoretical best practices in functional programming, highlighting how ChatGPT is used as an advisory tool rather than a direct source of patches. Ultimately, rather than replacing developer decision-making, ChatGPT functioned as a catalyst for discussion and refinement within the pull request workflow.

 \noindent \faCheckSquareO \textbf{Project Organization and Documentation Refinement:}
Beyond code modifications, developers frequently use ChatGPT to help improve project organization, improve documentation, and optimize workflow structures. This theme captures instances where ChatGPT’s suggestions influenced non-code contributions, such as file management and documentation strategies, rather than direct patch implementations.
From our dataset, we identified 6 of the 56 instances in which ChatGPT was consulted to streamline internal project structures. A representative case is pull request \#37 \cite{pr37} within the \texttt{Bananapus/nana-core} project (PN-2), where the developer sought an efficient way to manage file renaming within a pull request. The ChatGPT prompt specified the issue: \textit{``\ul{in a GitHub pull request, I have renamed a file. In the diff viewer, it shows one file deleted and another added, but I instead want to show the file as renamed.''}} ChatGPT suggested a structured file management approach to ensure that Git detected the renaming action instead of treating it as deletion and addition. The developer applied this suggestion, improving repository maintainability and commit traceability. No code patch was applied since the pull request focused on internal organization rather than functional modifications.

\begin{custombox}
{\faLightbulbO \hspace{0.01in} \textbf{Summary of RQ2 Results:}}
\textit{The analysis of 56 \PN instances identified six main reasons why ChatGPT-suggested patches were not directly integrated: adaptation to project needs, technical limitations, methodological guidance, functionality-related decisions, validation and refinement of implementation, and documentation improvements. Developers frequently adapted AI-generated suggestions rather than applying them verbatim, often modifying them to fit project-specific constraints or replacing them with more suitable alternatives. In some cases, patches were rejected due to system limitations or developer preferences. In addition, developers use ChatGPT to validate ideas, explore alternative approaches, or refine documentation and project structure rather than directly integrating its generated code. These findings indicate that while ChatGPT plays a role in decision-making within pull request workflows, its suggestions are typically subject to critical evaluation and transformation before adoption.}
\end{custombox}

\subsubsection{Implications for RQ2}
While RQ1 revealed that developers rarely integrate ChatGPT patches verbatim, our RQ2 analysis adds a crucial nuance: even when patches are not applied (\PN), ChatGPT influences developer decisions in meaningful ways. In these cases, developers assess suggestions critically and often extract conceptual guidance, architectural direction, or documentation support. This extends prior work~\cite{grewal2024analyzing, siddiq2024quality}, which focused on patch acceptance rates, by showing that AI-generated code can shape developer thinking and workflow even when it is ultimately not merged.

Our contribution also complements recent studies examining the limitations of generative AI in software engineering. While prior work has investigated correctness risks in AI-generated code~\cite{tanzil2024chatgpt}, cross-tool variability in patch quality~\cite{li2024copilot}, and the use of AI-generated metadata in PRs~\cite{xiao2024generative}, our findings offer a behavioral lens into how developers evaluate and selectively reuse AI suggestions. Additionally, we build on research into multi-hunk patch challenges~\cite{nashid2025characterizing} and automation limitations in code review~\cite{tufano202:IEEEPress} by illustrating the human-in-the-loop decision-making that occurs when AI-generated patches are assessed but not applied.

Recent evaluations have highlighted limitations in LLMs’ ability to reason about and adapt to project-specific constraints~\cite{moumoula2024crosslingual, ouedraogo2024unitgeneration, chen2024llmmobile}. These limitations likely contribute to patch rejection or modification. Our findings emphasize the need for AI systems that are better aligned with project context, coding conventions, and architectural patterns, an area that remains underexplored in current tooling.

\paragraph{\faLaptop \hspace{0.01in} For developers}
Developers use ChatGPT’s suggestions as conceptual starting points rather than direct solutions, often modifying or rejecting them based on project-specific requirements. This underscores the need for structured validation when using AI-assisted patches. Since ChatGPT-generated code may not always be directly compatible with existing architectures or coding styles, developers should critically assess AI suggestions within the context of system dependencies, maintainability goals, and pull request review feedback before applying them. This finding suggests that AI-assisted patch adoption can be improved by establishing clearer internal guidelines on when and how to incorporate ChatGPT’s code into ongoing development workflows.

Beyond direct code generation, ChatGPT influences how developers refine and document their implementation strategies. Our results show that even when ChatGPT’s patches are not applied, developers use AI-generated insights to improve documentation clarity, enforce project-specific conventions, and inform architectural decisions. This suggests that teams can benefit from integrating AI tools into pull request discussions, even when patches are not used directly, by treating them as a mechanism to refine implementation details and improve project documentation.

\paragraph{\faSearch \hspace{0.01in} For Researchers}
These findings highlight a crucial research gap to understand how AI-generated suggestions are evaluated, modified, or dismissed prior to pull request integration. Future research should analyze how developers iteratively refine AI-assisted patches, distinguishing between modifications that preserve AI-generated logic versus those that entirely rework AI suggestions. Studying decision patterns, such as why some ChatGPT-suggested patches are partially modified while others are rejected, could provide insights into when AI-generated solutions are most effective.

In addition, improving AI-generated patches requires more context-aware recommendation models. Current AI systems lack fine-grained awareness of project-specific constraints, repository structures, and architectural dependencies, leading to suggestions that are syntactically valid but not semantically useful. Future research could explore adaptive AI models that learn from previous pull request discussions and historical development patterns, ensuring that AI-suggested patches align with specific project norms and are more seamlessly integrated into pull request workflows. This refinement would help bridge the gap between AI-assisted coding and real-world software engineering practices.

\subsection{Results \& Discussion -- RQ3}
\noindent \textit{Under what circumstances does ChatGPT not generate patches in developer interactions (no existing patch-\NE), and how do developers use its responses under the self-admitted ChatGPT usage?} 
 
\begin{figure}[ht]
    \centering
    \includegraphics[width=\linewidth]{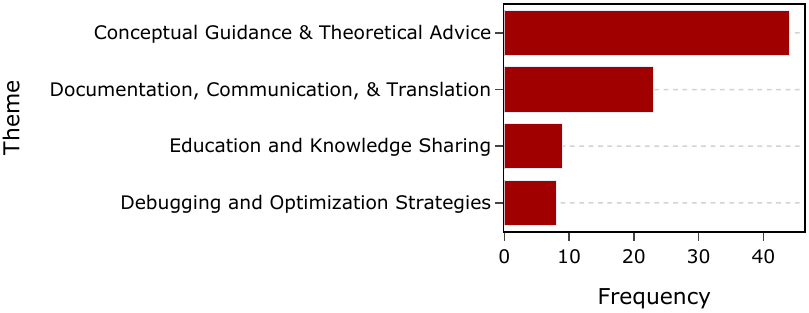}
    \caption{Showing the distribution of themes in the No patch generated class.}
    \label{fig:ne-class}
\end{figure}

As shown in Table~\ref{tab:sampling}, \tool classified 106 pull request instances as \NE. For our analysis, we manually validated and examined a representative sample of 84 instances, as described in Section~\ref{sec:patch-classification}. From the generated summaries of the 84 \NE instances, the employed card sorting analysis yielded four distinct themes, as illustrated in Figure \ref{fig:ne-class}. The identified themes are: Conceptual Guidance \& Theoretical Advice, Documentation, Communication \& Translation, Education and Knowledge Sharing, and Debugging \& Optimization Strategies.

\noindent \faCheckSquareO \textbf{Conceptual Guidance \& Theoretical Advice:} 
This was the most prevalent theme, with 44 of 84 instances where ChatGPT did not suggest patches but instead guided pull request decision-making. Developers refined code based on programming concepts, design principles and best practices, improving readability, maintainability, and overall design.
For example, in pull request \#1061 \cite{pr1061} within the \texttt{codecrafters-io/frontend} project (NE-36), a reviewer used ChatGPT to critique a variable’s naming convention, stating: \textit{``\ul{Bad naming, this should be something like selectedPricingFrequency. Reasoning: [ChatGPT link]''}}  The pull request author acknowledged the issue and noted: \textit{``\ul{[...] we don't seem to be getting better with naming [...] talk to ChatGPT about it [...] we need to pay extra attention to it.''}}  
Originally, the variable name \texttt{pricingFrequencyClicked} implied an action (clicking) rather than a state (selected frequency). After reviewing ChatGPT’s feedback, the developer renamed it to \texttt{selectedPricingFrequency}, improving clarity. While no code snippet was suggested, ChatGPT’s advice directly influenced the pull request decision, refining naming conventions.
The absence of a patch did not hinder decision-making but encouraged refinement. Reviewers used ChatGPT’s insights to justify changes, while maintainers ensured clarity before merging. These interactions shifted pull request discussions from issue resolution to proactive quality improvements, reinforcing AI’s role in better development practices.

\noindent \faCheckSquareO \textbf{Documentation, Communication \& Translation:}  
In 23 of the 84 instances, the discussions centered on refining the documentation, improving communication and ensuring accurate localization. Although no code snippets were generated, ChatGPT’s responses helped developers refine explanations, adjust terminology, and align documentation with project standards before merging changes.
A key example is pull request \#2230 \cite{pr2230} from the \texttt{faker-js/faker} project (NE-8), where a reviewer suggested re-writing a comment to be more positive and inclusive. The original comment stated:
\textit{\ul{``It would not be a good idea to have an English-only fallback list, because the Esperanto community is very diverse and international.''}} To improve the phrasing, the reviewer consulted ChatGPT, who suggested multiple re-wordings. The final version was collaboratively adjusted and implemented, with the pull request author noting:
\textit{\ul{``Thank you for your comment, I used this suggestion!''}}  
Although ChatGPT did not provide a code patch, its response influenced pull request decision-making by shaping documentation revisions before merging. AI-assisted refinements helped drive consensus in pull request reviews, leading to clearer and more inclusive communication.
Although ChatGPT did not generate a code patch, its response influenced pull request decision-making by reinspecting documentation before merging. AI-assisted suggestions helped reviewers justify edits and maintainers ensure clarity, shifting pull request discussions from minor wording adjustments to broader quality improvements.

\noindent \faCheckSquareO \textbf{Education and Knowledge Sharing:} 
This theme captures 9 of the 84 instances in which developers used ChatGPT to clarify programming concepts, system behaviors, or language-specific features. Instead of providing a direct patch, ChatGPT’s explanations helped developers make informed adjustments in pull requests by improving their understanding of technical constraints.  
A notable example is pull request \#2415 \cite{pr2415} from \texttt{digitalbitbox/bitbox-wallet-app} (NE-57), where a developer sought to determine the location of the Homebrew binary installation on a Mac M1 and how different commands influenced the installation paths:  
\textit{``On a Mac M1, \ul{what's the location of the brew binary?}''} 
ChatGPT clarified Homebrew installation differences, helping the developer determine the correct binary path. The pull request discussion validated this approach, with the reviewer finding the insight interesting and raising concerns about system-wide effects, such as \textit{\ul{“replace their normal brew [...] all future installs [...] wrong architecture.”}}  
While no patch was generated, ChatGPT’s response directly influenced the pull request by helping the developer refine installation logic. Understanding how Homebrew installs based on system architecture enabled a more informed decision on script modifications. AI-assisted clarification also guided pull request discussions, leading to further validation and refinement before merging.  

\noindent \faCheckSquareO \textbf{Debugging and Optimization Strategies:} 
We identified 8 of the 84 instances where ChatGPT assisted with debugging, performance optimization, and algorithm refinement, providing strategic insights without generating code snippets.  
An illustrative example is pull request \#5068 \cite{pr5068} from the \texttt{darklang/dark} project (NE-2), where a reviewer sought an alternative to improve the performance of using `vector` in C++, recalling: 
\textit{``\ul{I remember you mentioning that the performance of pushback isn't great. Do you recommend a different approach?''}}
The pull request author consulted ChatGPT for performance evaluation and alternative strategies. After reviewing the AI-generated insights, the author proposed a new approach:  
\textit{``\ul{[...] appending it to the front and then reversing should do it [...]}''}  
The developer switched to using \texttt{push} instead of \texttt{pushBack}, improving performance since inserting at the front typically has $O(1)$ time complexity, whereas \texttt{pushBack} could lead to $O(n)$. This discussion shaped the pull request decision, as ChatGPT provided optimization strategies rather than a direct coding solution.  
Although no patch was generated, ChatGPT influenced pull request decisions by helping the developer assess trade-offs and select a more efficient approach.  AI-assisted insights guided the pull request review, enabling an informed optimization decision before merging.

\begin{custombox}
\faLightbulbO \hspace{0.01in} \textbf{Summary of RQ3 Results}:  
\textit{The analysis of 84 \NE instances reveals that ChatGPT does not always generate patches but still influences pull request decision-making through conceptual guidance, documentation refinement, knowledge sharing, and debugging strategies. Developers used ChatGPT’s responses to improve variable naming, adjust documentation language, refine installation scripts, and explore performance optimizations. These non-patch responses shaped pull request discussions by providing explanations and alternative approaches, demonstrating that AI contributes to software development beyond code generation.}
\end{custombox}

These results build on and extend prior research on AI-assisted software development by highlighting a distinct usage pattern: developers often consult ChatGPT not for code generation but for conceptual refinement, documentation improvement, and knowledge support. While previous work has emphasized correctness limitations and patch quality issues in AI-generated code \cite{tanzil2024chatgpt, li2024copilot}, as well as the role of large language models in review automation \cite{tufano202:IEEEPress}, our findings point to a complementary function. They align with emerging views of large language models as collaborative assistants that support design thinking, aid code comprehension, and enhance communication, even in the absence of generated patches. These non-code uses suggest that ChatGPT contributes to software development in ways that extend beyond generation, offering new forms of value in pull request workflows that remain underexplored in prior literature.

\subsubsection{Implications for RQ3}  
\noindent Our findings highlight that ChatGPT frequently provides non-patch responses that still influence pull request decision-making. Developers do not always require direct patches; instead, they leverage AI-generated feedback for conceptual improvements, documentation refinements, knowledge expansion, and debugging strategies. These insights have important implications for both practitioners and researchers.  

\paragraph{\faLaptop \hspace{0.01in} Implications for Developers}  
ChatGPT’s non-patch responses help developers refine their code by offering structured feedback on best practices, naming conventions, and maintainability. AI-generated insights can support code reviews, enabling teams to assess design choices and enforce coding standards. In documentation workflows, ChatGPT improves clarity and consistency, allowing teams to incorporate AI-driven language refinements into their review processes. This improves the readability of the documentation and ensures technical accuracy before merging. For debugging and optimization, ChatGPT provides strategic recommendations rather than direct patches, helping developers identify bottlenecks, assess trade-offs, and refine performance strategies. Developers can use these insights to justify pull request decisions and explore alternative implementations before committing changes.  
By integrating ChatGPT into pull request workflows, developers can enhance collaborative decision-making, using AI not just for code generation but as a real-time knowledge assistant that strengthens software development practices.

\paragraph{\faSearch \hspace{0.01in} Implications for Researchers}  
Future research should explore how AI-generated non-patch responses impact long-term software design decisions. Investigating whether repeated exposure to AI-driven conceptual guidance influences coding best practices and architectural choices could provide insights into AI’s role in shaping developer decision-making over time.  
More studies could examine how AI explanations affect debugging efficiency at different experience levels. Research could assess whether AI-driven insights help junior developers bridge knowledge gaps or provide novel perspectives for experienced engineers in complex problem-solving scenarios.  
Furthermore, analyzing when and why developers trust AI-generated theoretical advice without direct patches could inform the design of AI systems that better align with developer expectations in collaborative software development.

\subsection{Results \& Discussion -- RQ4}
\noindent \textit{How does ChatGPT influence closed pull requests, and what factors contribute to these outcomes under the self-admitted ChatGPT usage?} 

\begin{figure}[h!]
    \centering
    \includegraphics[width=\linewidth]{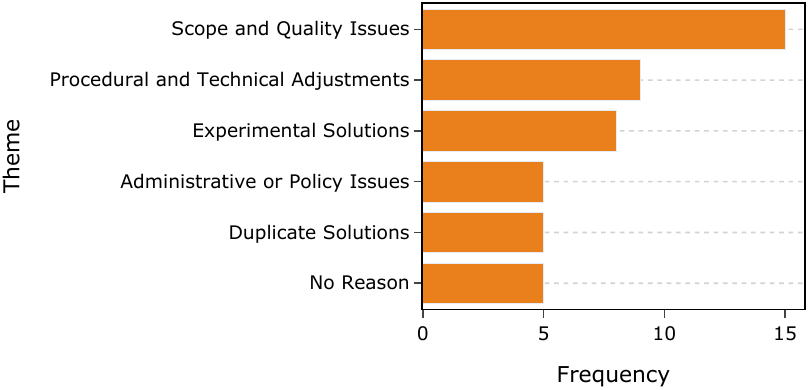}
    \caption{Showing the distribution of themes for instances where pull requests are closed without merging. }
    \label{fig:cl-class}
\end{figure}

We identified 53 instances of closed pull requests (\CL) for analysis. Following the same sampling methodology used in RQ2 and RQ3, we applied a 95\% confidence level with a 5\% margin of error, resulting in a stratified sample of 47 \CL instances for manual inspection. We then performed an in-depth qualitative analysis of these 47 cases, focusing on the interactions between developers and ChatGPT within the context of the pull request discussions. The qualitative analysis identified six key themes, as illustrated in Figure~\ref{fig:cl-class}. The most prevalent reason for pull request closure was attributed to Scope and Quality Issues, which accounted for 31.9\% of the cases. Other reasons included procedural and technical adjustments (19.1\%), experimental solutions (17.0\%), administrative or policy issues (10.6\%), duplicate solutions (10.6\%) and cases where no specific reason (10.6\%) was documented.

\noindent \faCheckSquareO \noindent\textbf{Scope and Quality Issues}  
This was the leading cause of pull request closures, accounting for 15 of 47 cases. Many pull requests were rejected due to quality concerns, workflow misalignment, or maintainability issues. Although ChatGPT’s optimizations were useful, developers struggled to integrate them into long-term architectural plans.
In pull request \#428~\citep{pr428} of the \texttt{gemini-hlsw/scheduler} project (CL-31), ChatGPT suggested replacing inefficient \texttt{list}-based membership checks with a \texttt{set}-based approach and explained Python’s \texttt{frozenset.issuperset()} behavior along with \texttt{\_\_contains\_\_} and \texttt{\_\_hash\_\_}. Although this improved efficiency, it did not resolve architectural issues related to object consistency during unpickling. 
Reviewers referred to ChatGPT’s responses to validate assumptions, but ultimately favored a fundamental architectural change over a band-aid fix. They emphasized the need for a long-term solution, stating:  
\textit{\ul{``This needs a proper fix rather than a band-aid solution; future developers will face the same issue if left unresolved [...]''}} and \textit{\ul{``[...] A generator should be used here. If we proceed with this approach (which raises red flags), it must be carefully revised [...]''}}.
ChatGPT’s optimization was valid, but the main issue was ensuring object consistency during unpickling. Rather than modifying the membership test, reviewers restructured unpickling by integrating it with \texttt{ResourceManager} and relocating it to the \texttt{lucupy} library.  
This case shows that, while ChatGPT-assisted optimizations can enhance performance, they may not align with architectural goals. The reviewers found that its insights were useful for validation, but prioritized structural changes for long-term sustainability.

\noindent \faCheckSquareO \textbf{Procedural and Technical Adjustments:} 
This theme encompasses pull requests requiring refinements or procedural modifications before integration. 9 of the 47 cases fell into this category, often involving technical restructuring, refactoring, or adherence to project workflows.
A representative case is pull request \#534 \cite{pr534} from \texttt{theosanderson/taxonium} - CL-47, which aimed to introduce a map visualization displaying sample origins in a phylogenetic tree. During the discussions, a reviewer explored the implementations using ChatGPT, which suggested leveraging \texttt{DeckGL} for rendering pie charts. The reviewer noted: \textit{\ul{``[...]This was ChatGPT's idea for that in \texttt{DeckGL}[...]''}}. The pull request author acknowledged the approach’s feasibility, stating: \textit{\ul{``[...]Neat! I think that approach could work. Let me see if I can get something working in a feature branch}[...]''}. 
Although viable, maintainers identified the need for significant back-end modifications, including restructuring metadata handling and creating new API routes. Due to these broader architectural changes, the pull request was closed in favor of continued development within the \texttt{add-map-view} feature branch. This case highlights how pull requests requiring substantial refinements are often redirected rather than merged immediately, aligning with trends in this category. ChatGPT played an exploratory role in shaping the implementation strategy, but the final decisions were guided by project-specific requirements.

\noindent \faCheckSquareO \textbf{Experimental Solutions:} 8 of the 47 closed pull requests were later replaced by more refined implementations. These pull requests were often \textit{experimental}, serving as \textit{proof-of-concept} solutions that were ultimately superseded by \textit{optimized alternatives}. In some cases, developers identified existing solutions that made ChatGPT-assisted contributions unnecessary for direct integration.
For example, pull requests \#2214~\citep{pr2214} and \#2212~\citep{pr2212} in the \texttt{open-learning-exchange/myplanet} repository (CL-29 and CL-36) introduced ChatGPT-generated French and Arabic translations for UI strings, addressing issues \#2216\footnote{\url{https://github.com/open-learning-exchange/myplanet/issues/2216}} and \#2219\footnote{\url{https://github.com/open-learning-exchange/myplanet/issues/2219}}. However, both pull requests were later closed, with their contributions consolidated into pull request \#2229\footnote{\url{https://github.com/open-learning-exchange/myplanet/pull/2229}}, which was merged.
This closure pattern aligns with the common practice of refining and merging multiple contributions into a single comprehensive pull request. Although ChatGPT facilitated the initial translation drafts, final refinements were likely developer-led to ensure linguistic accuracy and adherence to project standards. This underscores how ChatGPT-assisted contributions often act as stepping stones toward more polished, project-aligned solutions rather than directly leading to merged changes.

\noindent \faCheckSquareO \textbf{Administrative or Policy Issues:} five pull requests were closed due to administrative hurdles, including issues of the signing of the Contributor License Agreement (CLA), policy violations and housekeeping rules such as handling stale pull requests. A notable example is pull request \#468 \citep{pr468} in the \texttt{labdao/plex} project -- CL-5, where ChatGPT assisted in the development of a well-structured \texttt{README.md} for \texttt{Gnina} on the \texttt{PLEX} platform. Despite the quality of the documentation, the pull request was eventually closed after being flagged as stale. This finding highlights that ChatGPT's assistance does not always lead to pull request integration, as non-technical factors can prevent contributions from being merged, aligning with the goal of RQ to understand the role of AI in pull request closures.

\noindent \faCheckSquareO \textbf{Duplicate Solutions:}
pull requests in this category were closed because they duplicated existing efforts in the codebase or overlapped with other pull requests addressing the same issue. This includes cases where developers unknowingly submitted redundant pull requests or discovered an existing solution after submission. We identified five instances that reflect this theme.
For example, in pull request \#3792\cite{pr3792} of the \texttt{plausible/analytics} project - CL-20, the developer attempted to optimize how imported views are handled per visit by refining calculations in the \texttt{stats/base.ex} module. The primary concern was whether repeated expressions in queries to \textit{ClickHouse} led to redundant computations. ChatGPT was consulted to verify if \texttt{ClickHouse} recalculates identical expressions multiple times or reuses results for optimization. Based on the response of ChatGPT, the developer confirmed that \textit{ClickHouse} avoids redundant calculations, improving the efficiency of the query. However, pull request \#3792 was later closed in favor of pull request \#3830, which provided a cleaner and more efficient solution. The author of pull request \#3792 acknowledged the duplication, commenting: \textit{``[...] \ul{Closing in favor of \#3830.}''} and later stating: \textit{``[....]\ul{Uff, sorry for this} [...]. \ul{I had absolutely no intention of redoing your work **emoji** Unexpected coincidence **emoji**''}}.
This case highlights that while ChatGPT helped validate an approach, it did not prevent redundant work, as broader project awareness and coordination among contributors played a more decisive role in the pull request’s closure.

\noindent \faCheckSquareO \textbf{No Reason:} This theme captures 5 instances where the reason for pull requests closure was either not documented or remains ambiguous. Without clear explanations or comments from the project maintainers, the rationale for closing these pull requests is left open to interpretation. For instance, pull request \#2213\cite{pr2213} proposes the addition of Somali translations for various user interface (UI) strings within the \texttt{open-learning-exchange/myplanet} project - CL-9. The pull request aims to address issue \#2218 by incorporating localized Somali translations for menu items, error messages, buttons, and other textual elements within the application. The developer requested ChatGPT to translate UI strings from English to Somali. The input consisted of structured XML-like $<$string$>$ elements used in Android applications, where each text label has a corresponding translation. ChatGPT provided translations for a comprehensive set of strings. The developer shared these translations in the pull request, leveraging ChatGPT's ability to convert text into Somali while maintaining cultural and contextual relevance. However, the pull request was closed without merging with no explicit reason documented in the pull request discussion.

Prior studies have examined why pull requests are closed, highlighting factors such as contributor unfamiliarity, low code quality, and project maintenance policies~\cite{Tsay-2014, Gousios:ICSE:2016, soares2015acceptance}. Our findings complement this work by analyzing closures in the context of self-admitted ChatGPT usage. While some traditional reasons persist (e.g., scope misalignment, procedural issues), we identify new dynamics introduced by LLMs, such as AI-generated proofs-of-concept that are later superseded, or optimizations that, though valid, conflict with project-specific architectural norms. These patterns reveal emerging challenges in human–AI collaboration not addressed in earlier PR closure studies, and the persistent factors we observe align with established findings that PRs are often closed due to low code quality, misalignment with project goals, or procedural requirements such as policy compliance and workflow constraints~\cite{Tsay-2014, Gousios:ICSE:2016, soares2015acceptance}.

\begin{custombox}
\faLightbulbO \hspace{0.01in} \textbf{Summary of RQ4 Results:}
\textit{The analysis identified six primary reasons for the closure of pull requests involving ChatGPT: scope and quality issues (31.9\%), procedural and technical adjustments (19.1\%), experimental solutions (17.0\%), administrative or policy constraints (10.6\%), duplicate solutions (10.6\%), and undocumented reasons (10.6\%). The most common challenge stemmed from unresolved quality concerns and misaligned workflows, limiting the direct integration of ChatGPT-generated suggestions. Procedural and technical constraints frequently required restructuring or compliance with project workflows before acceptance. Additionally, many ChatGPT-assisted contributions served as experimental proofs-of-concept rather than final implementations. Other closures resulted from administrative hurdles, redundant contributions, or a lack of documented rationale. These findings highlight ChatGPT’s role in shaping pull request outcomes and the factors that influence its effectiveness in software development.}
\end{custombox}

\subsubsection{Implications for RQ4}
\noindent Our findings indicate that pull request closures involving ChatGPT-assisted contributions are frequently influenced by project alignment, maintainability, and procedural requirements. These insights have important implications for both practitioners and researchers, particularly in understanding how AI influences pull request outcomes and what factors contribute to closure decisions.

\paragraph{\faSearch \hspace{0.01in} For Developers}
\noindent While pull request closures occur in both AI-assisted and non-AI-assisted contributions, ChatGPT-generated patches introduce specific challenges due to their contextual misalignment and lack of repository awareness. To improve AI-assisted contributions and minimize pull request closures, developers should focus on the following key areas:

\begin{itemize}
\setlength{\itemsep}{0pt}
    \item \textit{Assess maintainability and project alignment}: Many AI-generated patches optimize performance, but do not fit long-term project goals. Developers should critically evaluate whether the suggestions of ChatGPT align with the architectural and maintainability requirements of the software before submission.
    \item \textit{Ensure compliance with repository policies}: pull requests were often closed due to missing procedural requirements, such as incorrect branch policies or lack of testing. Developers should manually verify that AI-generated patches conform to project guidelines.
    \item \textit{Check for redundant contributions}: AI-generated solutions sometimes duplicated existing efforts. Developers should review open pull requests and repository history before submitting an AI-assisted patch to prevent unnecessary duplication.
    \item \textit{Engage with maintainers early}: Some AI-assisted pull requests were closed without detailed feedback, indicating possible misalignment with the project goals. Developers should communicate with maintainers before submission to clarify expectations and increase the likelihood of acceptance.
\end{itemize}

\paragraph{\faSearch \hspace{0.01in} For Researchers}
Research should focus on addressing the specific challenges observed in AI-assisted pull request closures, particularly in areas where ChatGPT-generated contributions frequently fail. Key directions include:

\begin{itemize}
\setlength{\itemsep}{0pt}
\item \textit{Enhancing AI’s repository awareness}: ChatGPT lacks knowledge of project-specific guidelines, leading to pull request rejections due to non-compliance with repository policies. Research should explore techniques for integrating AI with real-time repository metadata to improve adherence to project standards.
\item \textit{Improving AI-generated code maintainability}: pull requests were frequently closed due to misalignment with the long-term project architecture. Future research should focus on adaptive AI models that analyze project history and architecture to generate more context-aware patches.
\end{itemize}

\subsection{Implications for Educators}
The findings of our study provide valuable insights for educators integrating generative AI tools such as ChatGPT into computer science curricula. By analyzing both merged and closed pull requests, along with varying patch integration levels, we highlight key learning opportunities. The classification into \PA, \PN, and \NE cases offers distinct educational value: \PA instances demonstrate AI-assisted coding in real-world, with varying integration percentages illustrating how developers selectively apply AI-generated code. \PN cases encourage students to critically assess AI suggestions, understanding why patches may be rejected due to maintenance concerns or project-specific constraints. \NE cases emphasize AI’s role beyond code generation, offering insights into conceptual learning, debugging strategies, and design discussions.

Additionally, our analysis of closed pull requests highlights the broader implications of AI-assisted contributions. Many pull requests were closed due to \textit{scope misalignment, maintainability concerns, or procedural issues}, demonstrating that AI-generated solutions must align with long-term project goals and organizational constraints. Educators can use these examples to teach students the importance of human oversight, team collaboration, and reviewer feedback when working with AI-assisted code. This perspective helps students develop a more comprehensive understanding of software engineering workflows beyond patch integration.

These insights are actively applied in \textbf{senior design projects at the University of Nevada Las Vegas (UNLV)}, where students analyze real-world instances of ChatGPT-assisted contributions. Case studies from this research are integrated into the curriculum, allowing students to engage in collaborative software development while exploring the ethical use of AI. ChatGPT is utilized for \textit{code review, debugging, refinement, and documentation}, aligning with industry practices.
Beyond senior design projects, these findings inform broader coursework: \PN cases help students develop critical evaluation skills in programming courses, while \NE cases illustrate how AI supports problem solving and iterative learning in software development. By studying both successful and rejected contributions, students gain a balanced perspective on the strengths and limitations of AI, preparing them for responsible AI-assisted development.

\subsection{Overall Findings and Discussion (RQ1 - RQ4)}
\label{sec:overall-discussion}
Our study analyzed 338 pull requests across four interaction categories: Patch Applied (\PA), Patch Not Applied (\PN), No Patch Generated (\NE), and Closed (\CL), to understand how ChatGPT influences collaborative software development workflows. Across these categories, ChatGPT-generated suggestions were used for a range of tasks, including bug fixing, small refactorings, configuration changes, documentation updates, and conceptual guidance without direct code adoption. Our analysis focuses on how developers evaluate and integrate such suggestions during pull request review, rather than on semantic correctness or long-term code quality. Accordingly, integration outcomes reflect contextual fit, integration effort, and maintainer trust, and rejection should not be interpreted as an objective assessment of code correctness.
In this study, we use the term ``impact'' to refer specifically to the influence of ChatGPT-generated suggestions on developer and maintainer decision-making during pull request review, rather than to any improvement in code quality, correctness, or performance.

Synthesizing findings across RQ1–RQ4, we identify several cross-cutting patterns that clarify how developers evaluate and adapt ChatGPT-generated suggestions in collaborative workflows:

\noindent \textbf{1. AI-generated patches are rarely adopted verbatim.} In \PA cases, the median integration rate was 25\%. Across 89 validated cases, developers most frequently engaged in \textit{structural integration}, \textit{selective extraction}, or \textit{iterative refinement}, reshaping ChatGPT’s code to meet project style, maintainability, and architectural constraints. High-integration cases were uncommon, highlighting the importance of context-aware adaptation.

\noindent \textbf{2. ChatGPT adds value even when patches are not applied.} In \PN cases, developers used ChatGPT as a conceptual starting point, adapting or rejecting its code based on project constraints and maintainability goals. Suggestions informed architectural decisions, validated alternatives, refined documentation, and improved organization, influencing workflows even without direct code integration.

\noindent \textbf{3. Non-code contributions shape development.} In \NE\ cases, ChatGPT influenced debugging strategies, naming conventions, design decisions, documentation quality, and knowledge sharing. These interactions demonstrate value beyond patch generation, supporting conceptual refinement and clearer communication.

\noindent \textbf{4. Pull request closures reflect both traditional and AI-specific barriers.} Consistent with prior findings on PR rejections~\cite{Tsay-2014, Gousios:ICSE:2016, soares2015acceptance}, many \CL\ cases were closed due to low code quality, scope misalignment, or policy non-compliance. Others reflected AI-specific patterns, such as proof-of-concept suggestions later superseded or optimizations conflicting with architectural constraints, illustrating new dynamics in human–AI collaboration.

\noindent \textbf{5. Developer judgment remains central.} Across categories, developers evaluated ChatGPT’s output against project requirements, coding standards, and architectural constraints. Suggestions were treated as one input among many, with final decisions shaped by reviewer feedback and domain expertise. AI-assisted contributions are therefore assessed within established review norms rather than granted special authority.

\vspace{5pt}

Viewed in the context of prior research on OSS pull request review~\cite{Tsay-2014, Gousios:ICSE:2016, soares2015acceptance}, these findings suggest that developers evaluate AI-generated code using largely the same criteria applied to human-authored contributions, including architectural fit, maintainability, and scope alignment. However, AI-assisted pull requests introduce additional negotiation around reuse: developers more frequently extract fragments, adapt ideas, or discard generated code while retaining conceptual guidance. Rather than exhibiting blanket distrust toward AI-generated code, maintainers treat it as a provisional artifact whose value depends on contextual integration effort. This extends prior PR review models by shifting emphasis from authorship credibility to adaptability and fit in AI-mediated contributions.

To further contextualize these findings, we performed a lightweight stratified analysis based on repository size, using the number of stars as a proxy (median = 96). Pull requests were divided into smaller and larger groups, and we examined whether integration outcomes differ across these contexts. While the distribution of PA, PN, NE, and CL cases showed moderate variation across groups, the overall integration patterns remained consistent. The median integration rate for PA cases was slightly higher in smaller repositories (22.13) than in larger ones (16.94), but this difference does not indicate a substantive shift in behavior. Qualitative behaviors were also consistent across both groups: in PA, developers engaged in structural integration, selective extraction, and iterative refinement; in PN, ChatGPT was used for adaptation and methodological guidance; in NE, it supported conceptual guidance, documentation, and debugging; and in CL, similar categories of rejection reasons were observed. These results suggest that the observed decision-making behaviors are not specific to a particular class of repositories but reflect broader characteristics of AI-assisted pull request workflows.

Although our dataset is limited to self-admitted ChatGPT usage, these decision-making patterns are not ChatGPT-specific. They reflect how developers evaluate externally generated code suggestions under uncertainty, an interaction increasingly common across agent-assisted programming tools such as IDE copilots and conversational assistants. In this sense, our findings characterize AI-mediated pull request review behavior more broadly, rather than the effects of a particular model or interface.

\vspace{5pt}
\noindent \textbf{Temporal scope and model evolution.}
Our dataset spans a period during which ChatGPT underwent multiple updates. While we do not explicitly track model versions, our analysis focuses on developer and maintainer responses to AI-generated suggestions during pull request review, rather than on the intrinsic quality of a specific model release. Across the DevGPT and extended datasets, we did not observe substantively different integration or rejection behaviors attributable to time period. This suggests that the decision-making patterns identified in this study reflect stable review practices in AI-assisted workflows, even as underlying models evolve.

\vspace{5pt}
\noindent \textbf{PatchTrack as a reusable research instrument.}
Beyond its use in this study, PatchTrack enables scalable analysis of how AI-generated code is integrated, adapted, or rejected within pull request workflows. Researchers can apply PatchTrack to examine integration behaviors across different repositories, programming languages, or AI tools, and to study how such behaviors evolve over time. It also supports comparative analyses of AI-assisted development practices across projects or ecosystems. While not intended as a developer-facing tool, PatchTrack may also offer practical insights into how AI-generated suggestions are incorporated within development processes, helping teams better understand integration patterns and review dynamics. In this sense, PatchTrack provides a reusable foundation for investigating AI-assisted software development workflows.

\vspace{5pt}
\noindent \textbf{Implications for practice.}
Practitioners can improve AI-assisted pull request workflows by treating ChatGPT-generated code as a provisional draft rather than a drop-in patch. This entails budgeting review effort for selective extraction, modification, or rejection of AI suggestions and encouraging brief disclosure of AI-assisted contributions during review. Such transparency helps distinguish exploratory assistance from deliberate design decisions and reduces ambiguity in collaborative evaluation.

More broadly, effective AI-assisted workflows reinforce existing review norms such as architectural fit and contextual reasoning rather than bypassing them. For tool builders, this highlights the value of systems that support contextual adaptation and explanation over verbatim generation. For researchers and educators, our findings underscore the importance of studying and teaching not only AI-based code generation but also critical review and adaptation in collaborative development.

\section{Related Work}
This section positions our study within three strands of research: pull request decision-making, AI-assisted software engineering, and clone detection techniques for patch comparison. Together, these areas provide the foundation for analyzing how ChatGPT-generated suggestions are evaluated and integrated in pull request workflows.

\subsection{Pull request decision-making:}
In collaborative software engineering and open-source projects, the decision-making process for pull requests is crucial to determine both the quality and the efficiency of the development. \citet{Gousios:ICSE:2014} and \citet{yu:2015} have examined how contributions are evaluated, highlighting the importance of peer review and social interactions. Various factors influence pull request acceptance, including technical merit, code quality, and alignment with project goals. For example, \citet{Tsay-2014} found that reviewers prioritize maintainability, readability, and adherence to coding guidelines when evaluating contributions. Similarly, \citet{Zhao:EMSE:2019} showed that pull requests that lack comprehensive tests are more likely to be rejected, reinforcing the importance of thorough validation. Complementing these findings, Soares et al.~\cite{soares2015acceptance} identified interaction history and test inclusion as strong predictors of pull request acceptance, laying a baseline for comparison with emerging AI-assisted workflows.

Beyond technical factors, social and ecosystem-wide dynamics also influence pull request decision-making. \citet{Dey:ESEM:2020} examined how technical attributes and the reputation of contributors affect the acceptance of pull requests in the NPM ecosystem, highlighting the importance of project constraints and community participation. Storey et al.~\cite{storey2016social} further emphasized that communication dynamics across different channels profoundly shape participatory cultures in software development, underscoring the collaborative challenges that arise in distributed workflows such as pull requests. In addition to social factors, prior research has highlighted the value of mining developer disclosures to better understand development practices. Yu et al.~\cite{yu2020jitterbug} introduced \texttt{Jitterbug}, a machine learning-based approach for detecting self-admitted technical debt (SATD) in code comments. Inspired by this idea, our study extends the concept of self-admitted artifacts to AI-assisted development by analyzing self-admitted ChatGPT usage in PR discussions.

Meanwhile, \citet{Azeem:ICSSP:2020} proposed a classification-based recommendation system to predict pull request actions (accept, respond, or reject), helping integrators make more informed decisions within pull-based workflows. Our study builds on this body of research~\cite{Ford:2019, golzadeh2019effect, legay2018impact, zhang2022pull} by extending the focus to AI-assisted pull request workflows, examining how ChatGPT influences pull request outcomes. By analyzing merged and closed pull requests, we provide empirical insights into how developers integrate, modify, or reject AI-generated patches, offering guidance on optimizing generative AI tools for improving pull request quality and accelerating decision-making.
Prior work characterizes how OSS developers evaluate human-authored pull requests, but largely assumes a human author. Our study extends this work by examining how established evaluation criteria are applied and adapted in AI-assisted pull request workflows.

\subsection{AI-Assisted Software Engineering:} 
Research on AI-assisted software engineering has examined the integration of ChatGPT-generated code in open-source projects, with studies such as Grewal et al.~\cite{grewal2024analyzing} and Siddiq et al.~\cite{siddiq2024quality} providing key insights. Grewal et al. analyzed 3,044 ChatGPT-generated code snippets in GitHub repositories, finding that 54\% of AI-generated code is retained in projects with minimal modifications. However, their study primarily measured retention rates and did not investigate the decision-making process behind the acceptance or rejection of the pull request. Similarly, Siddiq et al. evaluated the quality and security of ChatGPT-generated code, finding that AI-generated patches were rarely merged (only 12\%) due to missing documentation, undefined variables, and security vulnerabilities such as improper resource management and hard-coded credentials. Recent work by Ehsani et al.~\cite{ehsani2025promptgaps} highlights that gaps in developer prompts can lead to ineffective LLM outputs, further complicating the adoption of AI-generated code in collaborative workflows. Preliminary work by Latendresse et al.~\cite{latendresse2024librarian} assessed ChatGPT’s effectiveness in recommending software libraries, raising concerns about licensing issues and integration reliability; however, their focus remains on library usage rather than collaborative coding workflows. 
Our study builds on this work by examining why ChatGPT-generated patches are accepted, modified, or rejected in pull request workflows, and how these decisions differ from the evaluation of human-authored code.

Beyond studies on AI-generated code, prior work has investigated how automation influences developer workflows and decision-making. Tufano et al.\cite{tufano202:IEEEPress} found that automated code review tools, including those powered by LLMs, often ignore the human-in-the-loop context, limiting their reliability and developer trust. Cassee et al.\cite{cassee2020silenthelper, ghorbani2023autonomy} showed that automated systems such as Continuous Integration (CI) and GitHub bots alter code review behaviors and are critically evaluated by developers based on trust and perceived utility. Extending this line of inquiry, Shihab et al.\cite{shihab2024llmchatbots, shihab2023stalebot} analyzed the adoption of LLM-based chatbots and automation bots in pull request workflows, emphasizing challenges related to trust, accuracy, and context sensitivity. Consistent with these concerns, Tanzil et al.\cite{tanzil2024chatgpt} proposed a technique for detecting incorrect AI-generated code suggestions during code reviews, further underscoring the risks of hallucinated outputs.

In addition to studies on AI-generated code retention and quality, researchers have broadly examined AI-assisted software development, including the roles of ChatGPT and GitHub Copilot in code reviews, debugging, and software documentation~\cite{Jiang:2023,Guo:2024,Deng:2024,hao2024empirical,Vaithilingam:2022,Liang:2024,peng2023impact}. Although these studies demonstrate improvements in productivity and code quality, they often focus on controlled experiments rather than real-world open-source development~\cite{hou2023large,ju2023llama,siddiq2023exploring}. Recent work by Xiao et al.\cite{xiao2024generative} analyzed how developers edit and adopt AI-generated PR descriptions, identifying frequent partial reuse and frequent human intervention -- patterns echoed in our analysis of ChatGPT-suggested patches. Previous work has explored the role of AI in pull request workflow automation\cite{champa2024chatgpt, chouchen2024how}, but little research has examined how developers interact with ChatGPT in pull request discussions and adapt AI-generated patches to fit project-specific constraints. Our work addresses this gap by empirically analyzing ChatGPT’s impact in pull request workflows, particularly the decision-making processes that influence patch acceptance, modification, or rejection. This challenge is further complicated when dealing with multi-hunk patches, which often require semantic coordination and broader context management. Nashid et al.~\cite{nashid2025characterizing} recently showed that LLMs struggle with such coordination across hunk boundaries, a limitation also observed in our dataset.

Prior research on software evolution and API maintenance has also shown that developers often face challenges when integrating changes due to harmful API usage patterns and breaking changes. Ochoa et al.~\cite{ochoa2025harmfulapi} systematically characterized harmful API uses and available repair techniques, while Ochoa et al.~\cite{ochoa2022breakingbad} demonstrated how semantic versioning violations and API instability can complicate integration efforts. These challenges parallel the issues we observe when developers critically assess and adapt ChatGPT-suggested patches to fit evolving project architectures and avoid integration risks.

\subsection{Code clone detection:} \tool uses clone detection techniques to determine if a patch was applied, aligning our work with prior studies on clone detection~\cite{Hou:2019, Haibo:2021, pareco:2022, 6234404, kim:2017}. Baker~\cite{baker1995finding} introduced a system for detecting verbatim code copying through text-based clone detection, while \citet{kamiya2002ccfinder} developed \texttt{CCFinder}, a token-based tool that converts code into token sequences for similarity comparison—a method also employed in tools like \texttt{Sourcercc}~\cite{sourcerercc:2016}, \texttt{Rebedug}~\cite{6234404}, \texttt{PaReco}~\cite{pareco:2022}, and \texttt{VUDDY}~\cite{vuddy:2017}. \citet{baxter1998clone} demonstrated the use of abstract syntax trees (ASTs) for clone detection by identifying functional similarities despite syntax differences, as further emphasized by Lingxiao et al.’s \texttt{DECKARD}~\cite{Lingxiao:2007}. More recent studies leverage graph analysis and machine learning; for instance, Bowman et al.’s \texttt{VGRAPH}~\cite{vgraph:2020} uses code property triplets, while \citet{Siyue:2024} combine token-based and ML techniques, reporting improved performance over traditional approaches. 
Beyond general-purpose clone detection, research on framework-specific clone analysis has provided insights into software ecosystems. \citet{businge:2019} and \citet{kawuma:2016} studied clones in the Eclipse framework, reporting that clone occurrences range between 9\% and 10\%. These findings illustrate the prevalence of duplicated code even in structured development environments, reinforcing the importance of precise clone detection techniques, such as those implemented in \tool, to enhance software quality and maintainability.

\section{Threats to Validity and Trustworthiness}
\label{sec:threats}

\subsection{Threats to Validity}
For RQ1, the primary threat to validity lies in the accuracy of \tool. Reliance on n-gram tokenization with \( n=1 \) may introduce false positives, and normalization steps can obscure structural differences between ChatGPT-generated patches and applied code. Because \tool focuses on exact token overlap rather than semantic equivalence, substantially modified patches may be classified as \PN even when ChatGPT influenced the final implementation. 
To mitigate this threat, we manually validated a subset of cases and refined the classification procedure, though some edge cases remain.
Sensitivity analysis using the manually validated subset shows that the relative distribution of \PA, \PN, and \NE classifications and the median integration rate remain consistent with the full dataset, suggesting that the aggregate findings reported in RQ1 are robust to potential classification noise.

Another threat concerns dataset representativeness. Our analysis is limited to pull requests where developers explicitly shared ChatGPT-generated content, excluding private or undisclosed AI usage. This may introduce selection bias, as the dataset captures only visible instances of AI assistance. However, we view self-admitted usage as a conservative attribution strategy that prioritizes validity over coverage, and the observed decision-making patterns are not specific to ChatGPT but reflect how developers evaluate externally generated code suggestions in pull request workflows.

A related threat concerns temporal effects due to model evolution. Our dataset spans multiple ChatGPT updates, which may influence the nature of generated suggestions. However, because our analysis focuses on developer decision-making and integration behavior rather than model performance, and because we did not observe qualitatively different adoption patterns across time periods, we believe this does not threaten the validity of the reported findings.

A further limitation is the absence of a baseline comparison with non-AI-assisted pull requests. Such comparisons are challenging because AI usage is often undisclosed, making it difficult to construct reliable control samples. While historical pre-ChatGPT pull requests could serve as a proxy, temporal confounds such as changes in tooling, practices, and norms complicate interpretation. We therefore focus on self-admitted ChatGPT usage (SACU) and leave broader comparisons to future work.

Finally, a construct validity threat concerns the interpretation of ``impact.'' As clarified in our discussion, impact in this study refers to developer decision-making and integration outcomes during pull request review, rather than semantic correctness or long-term code quality. Acceptance, adaptation, or rejection of AI-generated suggestions represent the most immediate and observable effects of AI assistance in collaborative development, but they do not imply intrinsic code quality.

\subsection{Trustworthiness of Qualitative Findings (RQ2, RQ3, RQ4)}
For RQ2, RQ3, and RQ4, we assess the trustworthiness of our qualitative findings using Lincoln and Guba’s framework~\cite{lincoln1985naturalistic}.

\paragraph{Credibility}
We employed a rigorous qualitative analysis protocol based on the Framework Method and thematic extraction using card sorting~\cite{Gale2013, Spencer2009}. For RQ1, a sample of 230 pull requests (90 \PA, 56 \PN, and 84 \NE) was manually validated to assess the accuracy of PatchTrack’s automated classification. This process was conducted independently by the two authors using shared labeling criteria, achieving a Cohen’s Kappa of 0.85.

For RQ1–RQ4, we applied a multi-stage qualitative process involving one author, a research assistant, and a senior researcher. Structured summaries were independently created for overlapping subsets, consolidated through collaborative review, and coded into themes using card sorting. Disagreements were resolved through consensus meetings. All findings were grounded in developer conversations and pull request excerpts, with examples and artifacts included in the replication package~\cite{patchtrack:2025}.

\paragraph{Transferability}
Although derived from GitHub pull request discussions, our findings may transfer to other AI-assisted software development workflows. We provide detailed descriptions of the dataset, coding process, and themes to support contextual assessment.

\paragraph{Dependability}
We document all coding procedures, classification criteria, and iterative refinements to ensure consistency and support replication.

\paragraph{Confirmability}
To minimize researcher bias, thematic extraction followed structured decision rules and was cross-verified across independent coding rounds. Direct excerpts ensure interpretations remain grounded in the data.

\subsection{Limitations and Future Work}
One limitation of this study is the lack of explicit information about the ChatGPT versions used in the analyzed conversations. As model capabilities evolve, different versions may influence developer engagement and patch integration differently, but such distinctions could not be assessed with the available data.

Additionally, although our dataset spans multiple projects, we did not examine project-level factors such as repository activity or programming language, which may affect AI-assisted patch integration.

Our analysis is also limited to pull requests with self-admitted ChatGPT usage (SACU), excluding undisclosed or private AI interactions. This may introduce selection bias, as the dataset captures only visible AI assistance.

Future work should investigate how AI-generated recommendations can better incorporate project-specific constraints and maintainability considerations. Longitudinal studies examining the downstream evolution of ChatGPT-assisted patches and their effects on software quality and technical debt would complement our focus on developer decision-making and integration outcomes.

\section{Ethical Implications}
Ethical integrity is paramount in research utilizing public platforms like GitHub. While repositories are publicly accessible, privacy considerations remain crucial as our analysis involves examining pull requests without explicit consent. To address this, we ensured proper attribution by explicitly acknowledging all pull requests that contributed to our study and discussion. This includes citing relevant pull requests in our references and, where applicable, linking directly to their commits to appropriately credit the contributors.

To promote transparency and reproducibility, we have made our replication package publicly available, including datasets, analysis scripts, and supplementary documentation\cite{patchtrack:2025}. These measures uphold the ethical responsibilities of using open-source data while fostering reproducibility and openness in computer science research.

\section{Conclusion}

This study examined how developers engage with ChatGPT in pull request workflows by analyzing 338 real-world instances of self-admitted AI usage. By classifying cases into four categories (patch applied, patch not applied, no patch generated, and closed), we characterize how developers adapt, reject, or conceptually reuse AI-generated suggestions, revealing how AI assistance shapes decision-making in collaborative development.

Our analysis shows that full adoption of ChatGPT-generated patches is uncommon, with a median integration rate of 25\%. Qualitative analysis revealed recurring patterns of \textit{structural integration}, \textit{selective extraction}, and \textit{iterative refinement}, indicating that developers typically treat AI output as a starting point rather than a final solution. Even when no code is integrated, ChatGPT influences pull requests through conceptual guidance, documentation revisions, and debugging strategies. In closed pull requests, we observed both traditional barriers (e.g., code quality, scope misalignment) and AI-specific dynamics (e.g., discarded proofs-of-concept), underscoring the central role of developer judgment.

Consistent with our focus on developer decision-making, this study examines the immediate and observable effects of AI assistance during pull request review rather than semantic correctness or long-term code quality. In practice, our findings suggest that teams can more effectively incorporate AI assistance by treating AI-generated code as a provisional artifact, explicitly reviewing and adapting it for contextual and architectural fit rather than expecting direct reuse.

\backmatter

\section*{Declarations}
\subsection*{Funding}
This research was supported by the NEVADA NASA EPSCoR program under Award No. AWD2536/GR20392 and NSF Grant Award No.~\#2519136.
\subsection*{Ethics approval}
Not applicable
\subsection*{Informed consent}
Not applicable
\subsection*{Author contribution}
\textbf{John Businge}: Conceptualization, supervision, and review and editing of the manuscript.\\
\textbf{Daniel Ogenrwot}: Data collection, analysis, and writing of the original draft.

\subsection*{Data availability}
The datasets, analysis scripts, and supplementary documentation are publicly available at \href{https://zenodo.org/records/14978625}{https://zenodo.org/records/14978625}

\subsection*{Conflict of interest}
The authors declare that there are no known competing financial interests or personal relationships that could have appeared to influence the work reported in this paper.

\subsection*{Clinical trial number in the manuscript}
Not applicable

\bibliography{sn-bibliography}
\newpage

\appendix
\section{PatchTrack Design, Validation, and Robustness Analysis}
\label{appendix:patchtrack}

\subsection{PatchTrack Design and Implementation}

\texttt{PatchTrack} was adapted from \texttt{PaReco}~\cite{pareco:2022} to detect and classify ChatGPT-influenced patches in pull requests. While \texttt{PaReco} was designed for clone detection across divergent variant forks, it is not directly applicable to ChatGPT-assisted workflows. \texttt{PatchTrack} instead identifies ChatGPT-shared links, extracts code snippets from conversations, and compares them against pull request diffs.

Unlike \texttt{PaReco}, which relies on Bloom filters and can incur higher false positive rates, \texttt{PatchTrack} uses temporary hash tables for token matching. Hash tables offer average-case lookup time of $O(1)$~\cite{Cormen:2009, Bloom:1970}, enabling precise token-level matching with reduced false positives.

\subsection{Normalization and Language Support}

To ensure robust comparison across diverse codebases, \texttt{PatchTrack} applies normalization to both ChatGPT snippets and pull request diffs. This includes removing whitespace and comments, converting to lowercase, filtering non-ASCII characters, and applying syntax-aware heuristics based on file type. The tool currently supports 38 file types, covering all major languages in our dataset.

\subsection{Patch Classification Procedure}

Each pull request is analyzed at the hunk level. For each hunk, \texttt{PatchTrack} compares the n-gram tokens extracted from added lines (prefixed with `$+$`) to tokens in the ChatGPT snippet:

\begin{itemize}
\item If at least one match is found, the hunk is labeled \PA.
\item If no match is found but a ChatGPT snippet exists, the hunk is labeled \PN.
\item If no ChatGPT snippet exists, hunks are labeled \NE.
\item Unsupported file types are labeled \CC; processing failures are labeled \EE.
\end{itemize}

Pull request--level labels are derived from the aggregation of hunk-level classifications.

\subsection{N-Gram Sensitivity Analysis}

We evaluated \texttt{PatchTrack} under n-gram sizes $n = 1, 2, 3, 4$. Table~\ref{tab:ngram_comparison} reports the distribution of pull requests across \PA, \PN, and \NE categories for each configuration.

\begin{table}[ht!]
\renewcommand{\arraystretch}{1.2}
\small
\centering
\caption{Patch classification outcomes across different n-gram sizes based on 285 pull requests.}
\label{tab:ngram_comparison}
\begin{tabular}{lrrrr}
\toprule
\textbf{Classification Type} & \textbf{n = 1} & \textbf{n = 2} & \textbf{n = 3} & \textbf{n = 4} \\ 
\midrule
Patch Applied (\PA)     & 116 (40.7\%) & 83 (29.1\%)  & 47 (16.5\%)  & 38 (13.3\%) \\
Patch Not Applied (\PN) & 63  (22.1\%) & 96 (33.7\%)  & 132 (46.3\%) & 141 (49.5\%) \\
No Patch Exists (\NE)   & 106 (37.2\%) & 106 (37.2\%) & 106 (37.2\%) & 106 (37.2\%) \\
\bottomrule
\end{tabular}
\end{table}

Larger n-gram sizes reduced recall and increased \PN classifications due to stricter matching requirements, motivating the choice of $n=1$.

\subsection{Manual Validation and Performance Metrics}

We randomly sampled 230 pull requests (90 \PA, 56 \PN, 84 \NE) using a 95\% confidence level and 5\% margin of error. Each case was independently labeled by the two authors using shared criteria. Inter-rater agreement reached a Cohen’s Kappa of 0.85.

Table~\ref{tab:sampling} summarizes performance metrics across n-gram configurations.

\begin{table}[ht!]
\renewcommand{\arraystretch}{1.1}
\small
\centering
\caption{PatchTrack performance metrics (accuracy, precision, recall, F1) for each n-gram value.}
\label{tab:sampling}
\begin{tabular}{llrrrr}
\toprule
\textbf{N-Gram} & \textbf{Class} & \textbf{Acc.} & \textbf{Prec.} & \textbf{Recall} & \textbf{F1} \\ \midrule
\multirow{4}{*}{n = 1}
    & \PA & 97.2 & 96.6 & 95.6 & 96.1 \\
    & \PN & 95.4 & 92.9 & 94.5 & 93.7 \\
    & \NE & 100  & 100  & 100  & 100  \\
    & Overall & \textbf{97.5} & \textbf{96.5} & \textbf{96.7} & \textbf{96.6} \\ \hline
\multirow{4}{*}{n = 2}
    & \PA & 95.6 & 95.3 & 88.4 & 91.7 \\
    & \PN & 95.3 & 90.3 & 96.2 & 93.2 \\
    & \NE & 100  & 100  & 100  & 100  \\
    & Overall & 96.9 & 95.2 & 94.7 & 94.9 \\ \hline
\multirow{4}{*}{n = 3}
    & \PA & 95.9 & 88.1 & 88.1 & 88.1 \\
    & \PN & 95.9 & 94.9 & 94.9 & 94.9 \\
    & \NE & 100  & 100  & 100  & 100  \\
    & Overall & 97.2 & 94.3 & 94.3 & 94.3 \\ \hline
\multirow{4}{*}{n = 4}
    & \PA & 96.7 & 90.9 & 85.7 & 88.2 \\
    & \PN & 94.5 & 95.2 & 97.5 & 96.3 \\
    & \NE & 100  & 100  & 100  & 100  \\
    & Overall & 97.1 & 95.3 & 94.4 & 94.8 \\ \bottomrule
\end{tabular}
\end{table}

\subsection{Error Analysis and Robustness}

Manual inspection of misclassified cases under $n=1$ revealed seven errors: four false \PA and three false \PN classifications. Two primary error sources were identified:

\begin{itemize}
\item \textit{GitHub API limitations}: Diffs larger than 400 lines are truncated, leading to missed matches.
\item \textit{Single-line overlaps}: Short, syntactically generic lines occasionally triggered matches without semantic relevance.
\end{itemize}

To assess sensitivity to such errors, we recomputed the main RQ1 statistics using only the manually validated subset. The relative distribution of \PA, \PN, and \NE classifications and the median integration rate remained consistent with the full dataset, indicating that the reported aggregate trends are robust to classification noise.

\section{Qualitative Theme Definitions}
\label{app:themes}

This appendix provides the operational definitions and coding criteria for the six qualitative themes used to analyze ChatGPT’s contribution in pull request workflows (RQ1). These themes were applied during manual coding of validated \PA cases and reflect distinct modes of developer engagement with AI-generated output. Each pull request instance was assigned a single dominant theme based on the primary role ChatGPT played in the developer’s decision-making process.

\begin{table}[h]
\centering
\renewcommand{\arraystretch}{1.2}
\small
\begin{tabular}{p{3.5cm} p{11cm}}
\toprule
\textbf{Theme} & \textbf{Operational Definition} \\
\midrule
Constraint Handling &
ChatGPT generated code or suggestions that required modification due to project-specific constraints, such as API limitations, architectural rules, dependency requirements, or framework conventions. Developers selectively applied or reshaped AI output to comply with these constraints. \\

Direct Implementation &
ChatGPT-generated code was integrated with little or no modification. The suggested patch aligned closely with project requirements and coding standards, enabling near one-to-one reuse. \\

Iterative Refinement &
Developers used ChatGPT’s output as an initial draft and incrementally refined, corrected, or extended it through multiple revisions. This often involved reviewer feedback, testing, or additional design adjustments. \\

Knowledge Support &
ChatGPT contributed primarily through conceptual guidance rather than direct code integration. This includes explanations, documentation suggestions, design advice, or exploratory reasoning that informed developer decisions without being directly copied into the codebase. \\

Selective Extraction &
Developers extracted specific fragments or ideas from ChatGPT’s output (e.g., a function, logic pattern, or configuration snippet) while discarding the remainder. The AI served as a partial solution source rather than a complete implementation. \\

Structural Integration &
ChatGPT-generated code was adapted to fit within the project’s existing structure, APIs, or architectural patterns. This involved non-trivial restructuring to align with the surrounding codebase and system design. \\
\bottomrule
\end{tabular}
\caption{Operational definitions of qualitative themes used for coding ChatGPT’s contribution in pull requests.}
\label{tab:theme-definitions}
\end{table}

\appendix
\section{Qualitative Themes for RQ2--RQ4}
\label{app:themes-rq2-rq4}

This appendix provides concise operational definitions for the qualitative themes used to analyze pull requests where ChatGPT-generated patches were not applied (\PN), no patches were generated (\NE), or pull requests were closed (\CL). These themes characterize how ChatGPT influenced developer decision-making even when code was not directly integrated.

\begin{table}[h]
\centering
\renewcommand{\arraystretch}{1.15}
\small
\begin{tabular}{p{3.5cm} p{11cm}}
\toprule
\textbf{Theme} & \textbf{Operational Definition} \\
\midrule
Methodological Guidance &
ChatGPT influenced the developer’s approach, reasoning, or solution strategy without its suggestions being directly applied as code. \\

Internal Improvements and Documentation &
ChatGPT supported internal refactoring, documentation updates, or project maintenance activities that did not require direct patch reuse. \\

Specific Functionality Enhancement &
ChatGPT informed improvements to specific features or behaviors, but developers implemented customized solutions rather than applying the suggested patch. \\

Technical Limitations and Preferences &
AI-generated patches were rejected due to technical constraints, architectural incompatibilities, or a preference for alternative implementations. \\

Adaptation and Tailored Solutions &
Developers adapted ChatGPT’s conceptual advice to fit project-specific requirements, resulting in customized solutions rather than direct reuse. \\

Clarification and Correction &
ChatGPT helped clarify misunderstandings or correct reasoning, leading developers to revise their approach instead of applying new code. \\
\bottomrule
\end{tabular}
\caption{Qualitative themes for Patch Not Applied (\PN) cases.}
\label{tab:themes-pn}
\end{table}

\begin{table}[h]
\centering
\renewcommand{\arraystretch}{1.15}
\small
\begin{tabular}{p{3.5cm} p{11cm}}
\toprule
\textbf{Theme} & \textbf{Operational Definition} \\
\midrule
Conceptual Guidance and Theoretical Advice &
ChatGPT provided high-level guidance on design principles, programming concepts, optimization strategies, or naming practices without producing concrete code. \\

Documentation, Communication, and Translation &
ChatGPT assisted with documentation clarity, communication refinement, localization, or translation to improve accessibility and understanding. \\

Debugging and Optimization Strategies &
ChatGPT suggested debugging techniques, performance optimizations, or problem-solving strategies without supplying specific code patches. \\

Education and Knowledge Sharing &
ChatGPT was used to explain language features, code behavior, or concepts for learning and understanding rather than implementation. \\
\bottomrule
\end{tabular}
\caption{Qualitative themes for No Patch Generated (\NE) cases.}
\label{tab:themes-ne}
\end{table}

\begin{table}[h]
\centering
\renewcommand{\arraystretch}{1.15}
\small
\begin{tabular}{p{3.5cm} p{11cm}}
\toprule
\textbf{Theme} & \textbf{Operational Definition} \\
\midrule
Administrative or Policy Issues &
Pull requests closed due to administrative constraints, policy violations, contributor agreement issues, or repository housekeeping rules. \\

Experimental Solutions &
Proposed changes were exploratory or superseded by alternative implementations, making the pull request unnecessary. \\

Duplicate Solutions &
The pull request duplicated existing functionality or overlapped with another contribution addressing the same issue. \\

No Reason Stated &
The pull request was closed without an explicit explanation or with insufficient context to infer the reason. \\

Procedural and Technical Adjustments &
Closure resulted from unmet procedural requirements or unresolved technical refinements needed for acceptance. \\

Scope and Quality Issues &
Pull requests were closed due to excessive scope, quality concerns, workflow misalignment, or maintainability challenges. \\
\bottomrule
\end{tabular}
\caption{Qualitative themes for Closed pull requests (\CL).}
\label{tab:themes-cl}
\end{table}

\end{document}